\documentclass[aps,amsmath,amssymb,preprintnumbers,preprint,nofootinbib]{revtex4-1}
\pdfoutput=1
\usepackage{amsthm}
\usepackage{graphicx}
\usepackage{color}
\newcommand{\beq}{\begin{eqnarray}}
\newcommand{\eeq}{\end{eqnarray}}

\newcommand{\bmp}{\noindent\begin{minipage}{16cm}}
\newcommand{\emp}{\end{minipage}\vskip 7mm} 

\usepackage{dcolumn}
\usepackage{bm}
\usepackage{bbm}
\usepackage{subfigure}
\usepackage{pxfonts}
\usepackage{epstopdf}
\usepackage{slashed}

\newcommand{\vrel}{v_\textrm{rel}}

\newcommand{\be}{\begin{equation}}
\newcommand{\ee}{\end{equation}}
\newcommand{\bea}{\begin{eqnarray}}
\newcommand{\eea}{\end{eqnarray}}

\begin{document}

\begin{flushright}
CERN-PH-TH/2013-143
\end{flushright}

\vspace{0.5cm}
\begin{center}
{\Large\bf\color{black}  Gravity-mediated (or Composite ) Dark Matter  }\\
\bigskip\color{black}\vspace{1.0cm}{
{\bf Hyun Min Lee$^{1,*}$ , Myeonghun Park$^{2,\dagger}$ and Ver\'onica Sanz$^{3,4,\ddagger}$ }
\vspace{0.5cm}
} \\[8mm]

{\it $^1$Department of physics, Chung-Ang University, Seoul 156-756, Korea.  }\\
{\it $^2$Theory Division, Physics Department, CERN,  CH--1211 Geneva 23,  Switzerland.}\\
{\it $^3$Department of Physics Astronomy, York University, Toronto, ON M3J 1P3, Canada. } \\
{\it $^4$Department of Physics and Astronomy, University of Sussex, Brighton BN1 9QH, UK.  } \\
\end{center}
\bigskip
\centerline{\bf Abstract}
\begin{quote}
Dark matter could have an electroweak origin, yet communicate with the visible sector exclusively through gravitational interactions. In a set-up addressing the hierarchy problem, we propose a new dark matter scenario where gravitational mediators, arising from the compactification of extra-dimensions, are responsible for dark matter interactions and its relic abundance in the Universe. We write an explicit example of this mechanism in warped extra-dimensions and work out its constraints. We also develop a dual picture of the model, based on a four-dimensional scenario with partial compositeness. We show that Gravity-mediated Dark Matter is equivalent to a mechanism of generating viable dark matter scenarios in a strongly-coupled, near-conformal theory, such as in composite Higgs models.
\end{quote}

\vspace{0.5cm}

\begin{flushleft}
$^*$Email: hyun.min.lee@kias.re.kr \\
$^\dagger$Email: myeonghun.park@cern.ch  \\
$^\ddagger$Email: v.sanz@sussex.ac.uk
\end{flushleft}

\thispagestyle{empty}

\normalsize

\newpage

\setcounter{page}{1}

\section{Introduction}

In many extensions of the Standard Model (SM), such as Supersymmetry~\cite{susyDM} or Universal Extra-Dimensions~\cite{UEDDM}, Dark Matter (DM) relic abundance is obtained through electroweak interactions between the DM particle and the SM particles. Other extensions assume communication between DM and the SM through some kind of portal. For example, the Higgs portal~\cite{hportal} or Dark photons~\cite{darka}.

Yet the only property we are certain about DM is that it interacts gravitationally. In this paper we propose a mechanism to produce thermally the correct abundance of DM in the Universe, using exclusively gravitational interactions~\footnote{Non-thermal production of very heavy DM particles, or WIMPZILLAS, has been studied in Ref.~\cite{wimpzillas}.}. We will also focus on DM masses around the TeV scale, for reasons that will become clear in the next section. In this case, it is clear that four-dimensional gravity cannot annihilate enough DM particles. Instead, in Gravity-mediated Dark Matter (GMDM), the annihilation occurs through the exchange of gravity mediators. Gravity mediators are states around the scale of Dark Matter mass which arise via the compactification of extra-dimensions of space-time, namely the radion and massive graviton. 

A natural Gravity-mediation Dark Matter arises from warped extra-dimensions, and describing this model is the subject of the next section~\ref{xdims}.

Despite its name, Gravity-mediated Dark Matter is a scenario which has a dual description in terms of partial compositeness, where the strong sector is near-conformal. As we explain in Sec.~\ref{dualmodel}, the DM relic abundance computation is exactly the same in extra-dimensions as in composite models. The reason is that the relevant couplings of the dual of gravity mediators to SM is completely fixed by symmetries.

In Sec.~\ref{DMrelic} we describe the computation of DM relic abundance and the constraint it imposes on the DM and gravity mediator masses and on the scale of compactification. We extend the model, to account for bulk fermions in Sec.~\ref{pusht} and finish by discussing the collider constraints on the model, see Sec.~\ref{collider}.

\section{A model in warped extra-dimensions}\label{xdims}

\begin{figure}
\begin{minipage}{8in}
\hspace*{-0.7in}
\centerline{\includegraphics[height=8cm]{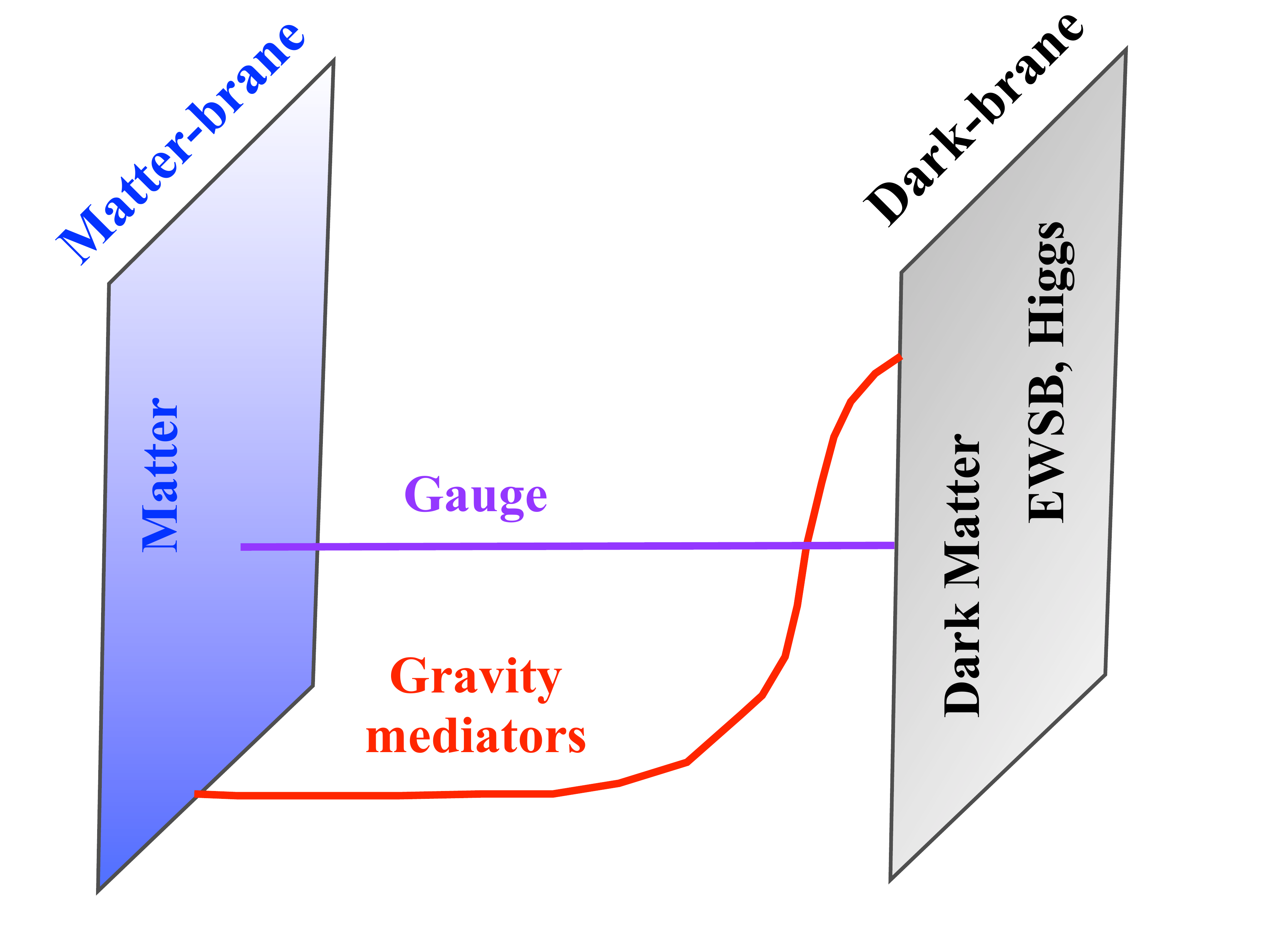}} \label{setupfig}
\hfill
\end{minipage}
\caption{
{\it
The set-up in extra-dimensions. Electroweak symmetry breaking, and the origin of Dark Matter stability and mass are related, hence their location on the same brane (IR brane, or Dark Brane). Gauge fields live in the bulk, and matter fields are located on the opposite brane (UV brane). EWSB is transmitted to fermions trough gauge and gravitational interactions.
} 
}
\label{setup}
\end{figure}

We now present the basic idea of Gravity-mediated Dark Matter (GMDM) in extra-dimensions. 
Let us consider the following class of five-dimensional (5D) metrics,
\begin{equation}
ds^2 = w(z)^2 \left(  \eta_{\mu\nu} dx^\mu dx^\nu - dz^2 \right),
\label{metric}
\end{equation}
where $z$ is the coordinate in the 5th dimension, and $w(z)$ is a smooth, decreasing or constant function of $z$. We are going to consider warped extra-dimensions, for reasons that will become clear later. A popular example of warping is Anti-deSitter (AdS) models, including Randall-Sundrum (RS)~\cite{RS}, is a particular case with $w(z)=1/(k z)$, where $k$ is the curvature of the five-dimensional (5D) space-time. The fifth dimension is compactified in an interval $z\in[z_0,z_1]$, and four-dimensional (4D) branes are located at both ends of the extra-dimension. Similar constructions could be obtained from a Klebanov-Strassler throat~\cite{KS}. We will denote the brane at $z_0$ the {\it Matter-brane} and the brane at $z_1$, {\it Dark-brane}, see Fig.~\ref{setupfig}.

Fields can be localized on branes, becoming truly 4D fields. But gravity and its excitations do necessarily propagate in the full 5D space-time. In our set-up, fields participating in electroweak symmetry breaking live on the Dark-brane, i.e. the Higgs $H$ and Dark Matter $X$. Dark Matter's mass and stability is linked to electroweak symmetry breaking, hence its localization on the same brane as $H$. Explicit realizations of this idea could be linked to, for example, a composite Higgs sector~\cite{CHM}, where $X$ could be part of the pseudo-goldstone sector and protected by a left-over symmetry~\cite{PomarolDM, chala}. In GMDM, however, we will not commit to specific realization of the dark matter sector and study scalar, vector and fermionic $X$. 

Gravity and gauge fields live in the bulk of the extra-dimensions, enjoying fully 5D dynamics, but their localization is different. Massless gauge bosons are de-localized in the bulk, with a flat profile. Gravity mediators (KK-graviton and radion) are peaked towards the Dark-brane as a result of the warping.

SM matter fields are localized on the {\it Matter-brane} although in Sec.~\ref{pusht} we will study the effect of pushing the top from the Matter-brane and into the bulk.

Instead of committing to a specific origin for $X$, we will describe its general properties: $X$ is a singlet under the SM, of mass at the electroweak scale, and stable due to a  quantum number conserved by the Dark-brane dynamics. As a singlet of the SM, $X$ interacts with the SM exclusively through gravitational interactions. The interaction of the massless graviton with any field is suppressed by $M_P$, and the leading interactions come from exchanging other gravitational fields, specifically the radion and the Kaluza-Klein (KK) massive gravitons, i.e. {\it gravity mediators}. In the following we describe how gravity mediators couple to the Matter and Dark sectors.

\subsection{Gravity mediators} \label{gravmed}
  
The graviton and the radion are described by the tensor and scalar fluctuations of
the metric, introduced as an expansion in Eq.(~\ref{metric}) 
\bea
d s^2= w(z)^2 \left( e^{-2 r} (\eta_{\mu \nu} + G_{\mu\nu} )- (1+ 2 r)^2 d z^2 \right) \ . \label{metric2}
\eea
where $G_{\mu\nu}$ and $r$ are 5D fields propagating in the extra-dimension. We are going to focus on the Kaluza-Klein (KK) resonance of the fields, including the effect of the whole tower.  In the following we denote  $G^{n}_{\mu\nu} (x,z)=G^{n}_{\mu\nu} (x) f^{n}_G(z)$  the n-th KK resonance of the graviton and $r(x,z)=r(x) f_r(z)$ the radion, and $f_{G,r}(z)$ are the wavefunctions. When we are discussing about a KK graviton, without specifying which excitation number, we will simply use the notation $G_{\mu\nu}$, dropping the $n$ label.

We consider the general interactions of a KK graviton $G_{\mu\nu}$ and the radion $r$ to a pair of  particles. The interaction arises by expanding the metric in Eq.~\ref{metric2} at linear order in $r$ and $G_{\mu\nu}$ in the action
\bea
{\cal S} \supset \int d^d x \sqrt{-g} {\cal L} \supset \int d^d x \sqrt{-g} \, w^2(z) \, \left( 2 r T - G_{\mu\nu} T^{\mu\nu}\right) 
\eea

Inserting the bulk profile of the fields and integrating out the extra-dimension, we obtain the 4D effective Lagrangian at dimension-five,
\bea
{\cal L}_{\rm KK}&=&-\frac{c^G_i}{\Lambda} G_{\mu\nu} \, T_i^{\mu\nu}    +\frac{c^r_i}{\sqrt{6} \Lambda} r \,  T_i \ , \label{dim-5}
\eea
where $T_i^{\mu\nu}$ is the energy-momentum tensor of species $i$, and is given in Appendix A, Eq.~\ref{fullTmunu}, and $T_i$ is its trace. $\Lambda$ is the compactification scale, related to the position of the Dark brane, $\Lambda=1/z_{Dark} \sim $ TeV. The coefficients $c_i^{r,G}$ arise by dimensional reduction from the 5D theory to the 4D low-energy effective theory. They are the overlap of wavefunctions of the fields in the bulk, e.g. $c^G_i\propto \int d z f_G(z) \, f_i(z)^2$. For a field $i$ localized on a brane at $z_*$, $f_i(z)^2 \propto \delta(z-z_*)$. For a field de-localized (flat) in the bulk, in our case massless gauge bosons, $f_i\propto $ constant. This can be easily seen by recalling that the equation of motion for the vawefunction of a massless spin-one field in the metric Eq.~\ref{metric} is given by $\partial_z (w(z) \partial_z f_V)=0$, and using the fact that the field has to satisfy Neumann boundary conditions on both branes~\cite{analogue}.

The KK gravitons $G_{\mu\nu}$ satisfies traceless and transverse conditions, $G_\mu^\mu=\partial_\mu G^{\mu\nu}=0$, which leads to a rather simple interactions,
\bea
{\cal L}_{\rm KK}
&=&-\frac{1}{\Lambda}G^{\mu\nu}\bigg[T^{\rm DM}_{\mu\nu} -c^G_V \,F_{\mu\lambda}F^\lambda\,_\nu  \nonumber \\
&&+ c^G_\psi\Big(\frac{i}{4}{\bar\psi}(\gamma_\mu D_\nu+\gamma_\nu D_\mu)\psi-\frac{i}{4}(D_\mu{\bar\psi}\gamma_\nu+D_\nu{\bar\psi}\gamma_\nu)\psi\Big)  \nonumber \\
&&+c^G_H\Big(D_\mu H^\dagger D_\nu H+D_\nu H^\dagger D_\mu H\Big)
\bigg].
\eea
with the traceless part of the energy-momentum tensor for dark matter (DM) given by
\bea
T^{\rm DM}_{\mu\nu}=\left\{ \begin{array}{lll} 
c^G_X \Big(-X_{\mu\lambda}X^\lambda\,_\nu+m^2_X X_\mu X_\nu\Big), \quad{\rm vector\,\,DM}, \\
c^G_\chi\Big(\frac{i}{4}{\bar\chi}(\gamma_\mu\partial_\nu+\gamma_\nu\partial_\mu)\chi-\frac{i}{4} (\partial_\mu{\bar\chi}\gamma_\nu+\partial_\nu{\bar\chi}\gamma_\nu)\chi\Big),\quad{\rm fermion\,\, DM}, \\
c^G_S\partial_\mu S \partial_\nu S, \quad\quad \quad{\rm scalar\,\,DM}. \end{array}\right.
\eea
Here,  $c^G_{X,\chi,S}, c^G_A, c^G_\psi, c^G_H$ are KK graviton couplings which are determined by the overlap between the wave functions of the KK graviton and fields in extra dimensions, see Ref.~\cite{RSbulk} for an example in AdS. $X(\chi,S)$,  $A$ , $H$ and $\psi$ denote the Dark Matter particle, gauge bosons, Higgs and SM matter fields, respectively.

The massless gauge fields do not contribute to the trace of the energy-momentum tensor at the tree level, but they generate trace anomalies at the loop level as
\be
T^\mu_{\mu,anom}=-\sum_a \frac{\beta_a(g_a)}{2g_a}\,F^a_{\mu\nu} F^{a\mu\nu}. \label{anom}
\ee
We note that including the linear radion couplings, non-derivative radion interactions to massive scalar and vector particles are fixed by dilatation symmetry to
\bea
{\cal L}_{\rm non-deriv}&=&-\bigg(\frac{r}{\sqrt{6}\Lambda}-\frac{r^2}{6\Lambda^2}\bigg)\Big(c^r_H m^2_A A_\mu A^\mu+c^r_X m^2_X X_\mu X^\mu\Big) \nonumber \\
&&+2\bigg(\frac{r}{\sqrt{6}\Lambda}
-\frac{r^2}{3\Lambda^2}\bigg)\Big(c^r_H m^2_h h^2+c^r_S m^2_S S^2 \Big ) + \frac{r}{\sqrt{6}\Lambda} m_{\psi} \bar{\psi} \psi
\eea
where use is made of gauge boson and real scalar mass terms as $\frac{1}{2}m^2_A\, e^{-\sqrt{\frac{2}{3}}\frac{r}{\Lambda}} A_\mu A^\mu $ and $-\frac{1}{2}m^2_S \,e^{-2\sqrt{\frac{2}{3}}\frac{r}{\Lambda}} S^2$, respectively \cite{csaki}. 
For comparison, in the dilaton case where dilatation symmetry is not extended to gravity \cite{dilaton}, there is no distinction between gauge bosons and scalars, and dimensionful parameters after electroweak symmetry breaking are replaced by a dilaton factor, $f\,e^{\sigma/f}$, where $f$ is the scale symmetry breaking scale and $\sigma$ is the dilaton.  Thus, choosing a canonical dilaton field as ${\bar\chi}=\chi-f$, the dilaton couplings to a pair of bosons are proportional to $2{\bar\chi}/f+{\bar\chi}^2/f^2$ \cite{dilaton}, so the quadratic dilaton couplings are different from the radion case.
Here, we ignored the mixing between the radion and the Higgs.
As will be shown later, the quartic couplings between the radion and the massive bosons will be important for calculating the relic density for bosonic dark matter with the radion mediator.

In warped extra-dimensions, there is a hierarchy of couplings of the graviton to Dark-brane, bulk and Matter-brane, respectively. Indeed, in the setup of Fig.~\ref{setupfig}, one obtains~\cite{gravitonus,doubleprod} 
\bea
& & \textrm{\bf Dark-brane fields : } c^G_X \simeq c^G_{H} \simeq {\cal O}(1)  \ , \\
& &  \nonumber \\
& & \textrm{\bf Bulk fields : } c^G_A \simeq \frac{1}{\int_{Dark}^{Matter} w(z) \, d z}   \ ,\\
& & \textrm{\bf Matter-brane fields : } c^G_{\psi} = \left(\frac{z_{Matter}}{z_{Dark}}\right)^{\alpha} \ ,
\label{cs}
\eea
where $\alpha > 1$. In AdS models, the value of $c^G_{\gamma,g}$ is
\bea
c^G_{\gamma, g}=2 \frac{1-J_0(x_G)}{\log\left(\frac{M_{Pl}}{TeV}\right) \,  x_G^2 \, |J_2(x_G)|} \label{cgamma}
\eea
where $x_G=3.83$ is the   the first zero of the Bessel function $J_1$, although localized kinetic terms could change the value of $x_G$~\cite{loc-kin}. Here we see explicitly the suppression by $\int w(z) d z = \log(M_P/TeV)\simeq {\cal O}$(0.03). Note that in this expression we have neglected the effect of higher KK-resonances, keeping only the lightest one.

The wavefunction of the graviton is more peaked towards the Dark-brane than that of the radion. This leads to differences in the degree of hierarchy among the couplings, but one still finds $c^r_{H,X} \gg c^r_{\psi}$. Moreover, the tree-level coupling to massless gauge bosons vanishes ($T_{\mu}^{\mu}$ is zero at tree-level) but loop-induced effects would generate this coupling, see Eq.~\ref{anom}~\footnote{See Ref.~\cite{real-radion} for a detailed calculation of these effects in AdS.}. Therefore, the radion couplings are characterized by
\bea
c^r_{H,X} \gg c^r_A \, \ , c^r_{\psi} \ .
\eea

Regarding the masses of the KK-graviton, it is $m_G \lesssim \Lambda$ and the exact relation depends on the metric. In AdS models the mass of the KK-graviton is related to $k$, Planck mass ${\bar M}_{P}$ and $\Lambda$ by
\bea
m_G =  \frac{k}{M_{Pl}}  x_G \Lambda
\eea
where one expects $k \lesssim M_{Pl}$. In other metrics, the relation between the curvature and the graviton mass would be different but, generally speaking, one expects a healthy theory to satisfy $m_G \lesssim \Lambda$. $\Lambda$ in Eq.~\ref{dim-5} is the scale suppressing dimension-five operators involving a gravity mediator and two other particles, hence expected to be larger than the mass of the fields we consider in the effective theory.

On the other hand, the radion mass is a model-dependent parameter, related to the mechanism of stabilization of the extra-dimension, as in absence of stabilization the radion is exactly massless. For example, in RS models one could assume the Goldberger-Wise mechanim~\cite{GW}, and in this framework the radion mass is a function of the vacuum expectation value and mass of the stabilizing field~\cite{GW-pheno}.

   
\section{Dual model}\label{dualmodel}

Extensive research on the applications based on the AdS/CFT correspondence~\cite{AdSCFTgen}, points at a duality between strongly coupled theories in $D$ dimensions and a gravitational dual in $D+1$ dimensions, even beyond supersymmetric or exactly conformal theories~\cite{LisaPoratti}.  This holographic duality is often a qualitative statement between strongly coupled systems (the target theory) and an {\it analogue computer}~\cite{analogue}, a theory on higher dimensions with improved calculability. We dedicate this section to describe the holographic dual of the model we presented in the previous section, whose schematic representation is given in Fig.~\ref{setupfig}.   

\begin{figure}
\begin{minipage}{8in}
\hspace*{-1.0in}
\centerline{\includegraphics[height=8cm]{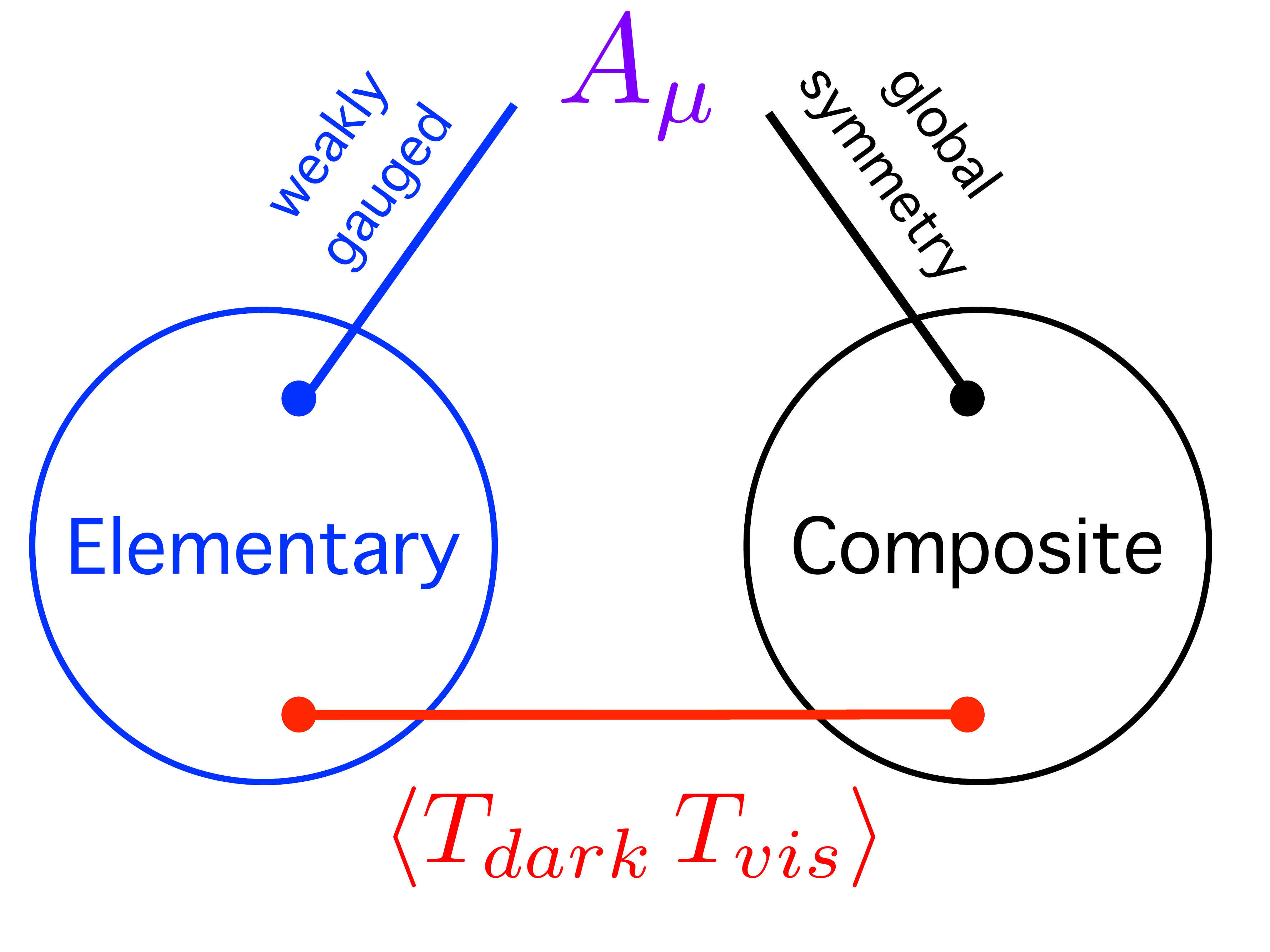}} \label{dualfig}
\hfill
\end{minipage}
\caption{
{\it
Dual picture
} 
}
\label{setup}
\end{figure}

The dual picture is portrayed in Fig.~\ref{dualfig}. The bulk of the extra-dimension encodes the RG evolution of the 4D Lagrangian, with the Matter-brane and Dark-brane representing the UV and IR boundary condition of the running, respectively. As one moves from the Matter to the Dark branes, the effect is one of integrating out degrees of freedom. At a position $z_*$ the {\it local} cutoff is related to the UV cutoff as~\cite{lisa-matt}
\bea
\Lambda (z_*) = \omega(z_*) \Lambda_{UV} \ . 
\eea

The running stops at the Dark brane: the presence of the Dark brane is signalling that a sector of the theory is undergoing confinement, and as a result {\it composite } states, the Kaluza-Klein modes, appear at low energies, hence the localization towards the Dark brane. Fields localized near or on the Matter brane do not strongly participate on the strong dynamics encoded near the Dark brane, and are then called {\it elementary}. Localization towards a brane is then the equivalent to the degree of compositeness of the field.   

De-localized (flat) gauge fields in the extra-dimension represent global symmetries of the composite sector, which are weakly gauged by the UV dynamics~\cite{csaba-fat}. They are therefore a mixture of composite and elementary field, much the same as the $\rho-\gamma$ mixing in QCD~\cite{ami-qcd,alex-qcd,QCD-hol}.

 Gravity mediators do also have an interpretation from the dual point of view. Their presence is a manifestation of a conformal symmetry of the composite sector, which is spontaneously broken by the strong (composite) dynamics.

 The radion is dual to the goldstone boson from dilatation symmetry in 4D~\cite{real-radion,Witek-Jiji}, the {\it dilaton} $\tilde{r}$. As such, couplings will arise of the form
 \bea
 \frac{\tilde{r}}{\Lambda_{\tilde{r}}} \partial_{\mu} J^{\mu}
 \eea
 where $J^{\mu}$ is the global current whose spontaneous breaking at the scale $\Lambda_{\tilde{r}}$ leads to the emergence of the Goldstone boson $\tilde{r}$. In general, the global current is given by
 \bea
 J^{\mu}= T^{\mu\nu} v_{\nu} \ ,
 \eea
 and if the symmetry is dilatation symmetry, $v_{\nu}$ has the form
 \bea
 v_{\nu} = \lambda x_{\nu}
 \eea
leading to a coupling to trace of the stress tensor ($T$),
 \bea
- \frac{c_i}{\Lambda_{\tilde{r}}} \tilde{r} \, T_i 
 \eea
 where $\Lambda_{\tilde{r}}$ is the symmetry breaking scale, $\Lambda_{\tilde{r}} \simeq \Lambda=1/z_{Dark}$. $c_i$ encodes the degree of compositeness of species $i$, with a large value indicating a large mixture with the composite sector. Similarly, the coupling of the dilaton to massless gauge bosons will follow the same structure as in Eq.~\ref{anom}. 
 
 The dual interpretation of the massive graviton is not so well understood, although some work has been done to link to the 4D $f_2$ resonance in QCD to a KK-graviton in AdS~\cite{amif2}. 
 
 We interpret the massive KK graviton is the manifestation of a CFT diffeomorphism invariance, broken spontaneously by the Dark Brane.
The massless spin-two field $\theta^{\mu\nu}$ is conserved, $\partial_{\mu} \theta^{\mu\nu}=0$ in the absence of breaking. As in the radion case, it couples to a conserved current $\partial_{\mu} J^{\mu}=0$. The massive spin-two corresponds to the breaking of this diffeomorphism invariance by $\partial_{\mu} \theta^{\mu\nu}=a^{\nu}$. The operator $a^{\mu}$ corresponds to a massive vector field, which is {\it eaten} by the massless spin-two field~\cite{massives2}. When joining together, the spin-two massless field and the massive vector will lead to a massive spin-two state $\tilde{G}$, the dual KK graviton. As long as the composite sector preserves Lorentz, gauge and CP invariance, the coupling of the massive spin-two resonance to two other particles will be given by~\cite{gravitonus}
 \bea
- \frac{c_i}{\Lambda_{\tilde{G}}} \, \tilde{G}_{\mu\nu} \, T^{\mu\nu}_i \ ,
 \eea
 where $\Lambda_{\tilde{G}}\simeq \Lambda$, which follows exactly the form of Eq.~\ref{dim-5}.  
 
 In summary, the dual picture of our warped extra-dimensional model is a model of partial compositeness, where the gravity mediators are composite states manifesting a broken conformal  symmetry in 4D at a scale $\Lambda$. Dark-brane states are fully composite states, whereas Matter-brane states correspond to elementary states. Bulk gauge fields are partly composite, with the gauge bosons coming from weakly gauging part of the global symmetry in the composite sector. Gravity mediators (radion/KK-gravitons) are resonances whose properties manifest the breaking of conformal invariance by the strong dynamics.

\section{Dark Matter relic density calculation}\label{DMrelic}

As the dark matter particle is assumed to be a singlet of the SM, all couplings with the SM occur through a graviton or radion  exchange. In Sec.~\ref{gravmed}, we discussed the hierarchy among the coupling of the radion and KK-gravitons to different species. Specifically,
\bea
c_{H,X} \simeq {\cal O} (1) \gg c_V \gg c_{\psi} \ .
\eea
Due to this hierarchy, we will focus on the gravity mediation processes $X \bar{X} \to H H^{\dagger}$.

Note that we will consider the exchange of the whole KK tower obtaining a compact expression in terms of the metric, as explained in Ref.~\cite{SRsa}. As dark matter is a singlet and resides on the IR brane where the Higgs boson is localized, it could also annihilate into a pair of the SM particles through the Higgs boson exchange by Higgs portal interactions. In particular, scalar dark matter can have an extra unknown dimensionless coupling to the Higgs boson.  
In our discussion, we will assume that Higgs portal couplings are subdominant and comment on their effect on dark matter annihilations.

\subsection{KK graviton mediators}
We will focus on the processes involving Dark-brane fields as their coupling is the largest. In other words, we are going to focus on DM annihilation into Higgs degrees of freedom,
\bea
X \, \bar{X} \to H \, H^{\dagger}
\eea
where a graviton/radion is exchanged.

Note that $H$ can be expressed as,
\begin{equation}
H = \frac{1}{\sqrt{2}}
\left( 
\begin{array}{c} 
\phi_1+ i \phi_2 \\ 
\langle h \rangle + h + i \phi_3 
\end{array} 
\right)
\end{equation}
where $\langle h \rangle$ is the vacuum expectation value of the neutral part of the doublet, and $h$ is the physical Higgs boson after EWSB. One can view $\phi_{1,2,3}$ as being {\it eaten} by the $W^{\pm}$ and $Z$ fields, providing the longitudinal polarization of a massive gauge boson. But in this set-up, $\phi_{1,2,3}$ and the massless $W^{1,2,3}$ bosons have different localizations, hence different couplings to $X$ and the gravity fields. We will then use the notation of $W_{L}$, $Z_{L}$ to denote the longitudinal fields, members of the Dark-brane field $H$. Similarly, we will denote by $W_{T}$, $Z_{T}$, the transverse part of the fields, which are bulk fields.

To compute the relic abundance of Dark Matter $X$, one needs to obtain the matrix element involving graviton exchange,
\bea
{\cal M} = c^G_{X} \, c^G_\phi \, T_{\mu\nu}^{X} \, P^{\mu\nu,\mu'\nu'} \, T_{\mu'\nu'}^{SM}
\eea
where the propagator $P^{\mu\nu,\mu'\nu'}$ is written in Appendix A, Eq.~\ref{PropG}.

The end result is quite transparent, and depends on the spin of the DM particle $X$.
In general, one can express the Dark Matter annihilation cross section as
\bea
(\sigma v)_{X_sX_s \rightarrow \phi \phi}  =  \frac{(c^G_X c^G_\phi)^2}{\Lambda^4} \, \, \frac{(a_s+b_s v^2+c_s v^4) \, m^6_X}{(4m^2_X-m^2_G)^2+\Gamma^2_G m^2_G}   \label{xstf}
\eea
where $G$ is the graviton field and $\phi$ denotes the Higgs boson $h$ and $Z,W$ gauge bosons, and we have neglected terms of the order ${\cal O} (m_{\phi}/m_X)^2$. See Appendix B for details.

The width of the graviton can be written as
\bea
\frac{\Gamma_G}{m_G} = \frac{1}{240 \pi} \left( \frac{m_G}{\Lambda} \right)^2 \label{GammG} 
\eea
in the limit of $m_G \gg m_{h, Z, W}$ (see Appendix B Eqs.~\ref{gammahh} for details). One can then safely neglect width effects for the heavy graviton case. Moreover, if $m_X \gg m_{\phi}, m_{G}$, and $c_H \gg c_V$, the annihilation cross section simplifies to 
\bea
(\sigma v)_{SS \rightarrow \phi \phi} &\simeq& \frac{ 3(c^G_S c^G_H)^2  }{ 16\pi} \,\frac{m^2_S}{\Lambda^4}\left(\frac{m^4_Z}{m^4_G}+2\frac{m^4_W}{m^4_G}\right), \\
(\sigma v)_{\chi{\bar\chi} \rightarrow \phi \phi} &\simeq& \frac{ (c^G_\chi c^G_H)^2v^2  }{ 576\pi} \,\frac{m^2_\chi}{\Lambda^4 } \ , \\
(\sigma v)_{XX \rightarrow \phi \phi} &\simeq& \frac{ (c^G_X c^G_H)^2 }{ 54\pi}\frac{ m_X^2}{\Lambda^4}  \ .
\eea
Therefore, as summarized in Table I, both scalar and vector dark matters annihilate as an s-wave while fermion dark matter is a p-wave suppressed. 
We note that when the longitudinal and transverse components of a massive gauge boson have the same coupling to dark matters, $c^G_H=c^G_V$,
the annihilation cross section for scalar dark matter becomes proportional to $v^4$, i.e. d-wave, as shown in Appendix B. This would be the case in the original RS model where the electroweak gauge bosons are localized on the TeV brane. However, in this case, other annihilation channels into massless gauge bosons and fermions would equally contribute.

\begin{table}[h!] 
\setlength{\tabcolsep}{5pt}
\center
\begin{tabular}{|l |c |c| c|} 
\hline \hline 
Mediator & X (s=0) & X (s=1/2) & X (s=1)
\\
\hline
 Graviton & s-wave & p-wave  & s-wave \\
Radion &  s-wave & p-wave &  s-wave \\
\hline \hline
\end{tabular}
\caption{\it Suppression in Dark Matter annihilation to Standard Model particles as a function of the Dark Matter spin and type of mediator.}
\label{table:collbounds} \vspace{-0.35cm}
\end{table}

We note that there could be tree-level Higgs portal couplings \cite{hportal} such as $\lambda_S S^2|H|^2$, $\lambda_\chi {\bar\chi} \chi |H|^2/\Lambda$ and $\lambda_X  X_\mu X^\mu |H|^2$ on the IR brane, for scalar, fermion and vector dark matter, respectively. The first coupling $\lambda_S$ is a renormalizable dimensionless parameter, while the latter two couplings, $\lambda_\chi, \lambda_X$ are non-renormalizable couplings which depend on a UV completion. 
We focus on the Higgs portal coupling for scalar dark matter but the discussion applies similarly to dark matter of other spins. First,  below WW threshold, $\gamma\gamma, gg$ and $f {\bar f}$ channels with KK graviton mediation are d-wave suppressed while
the $f{\bar f}$ channel with Higgs mediation is s-wave. But, the Higgs portal coupling could not be dominant in determining the relic density except near the resonance, due to the stringent XENON100 bounds \cite{hportal-xenon}. 
Second, above WW threshold,  for which s-wave WW/ZZ/hh channels with Higgs mediation are accessible, we should compare between them and the corresponding channels with KK graviton mediation.
In this case, for $|\lambda_S |\ll  c_S c_H m^2_S / \Lambda^2$, the effect of the Higgs portal term can be suppressed. Henceforth, we assume that the KK gravitons give rise to dominant contributions to the annihilation cross sections of dark matter. 

From the thermal average cross section,
\be
\langle\sigma v\rangle=a+ b v^2,
\ee
with $v^2=6/x_F$ where the freezeout temperature gives $x_F=m_S/T_F\simeq 20$, the relic density is determined by
\bea
\Omega_{\rm DM} h^2=\frac{2.09\times10^{8}\,{\rm GeV}^{-1}}{M_P\sqrt{g_{*s}(x_F)}(a/x_F+3b/x^2_F)}.
\eea

\begin{figure}[t]
\centering
\includegraphics[width=6cm]{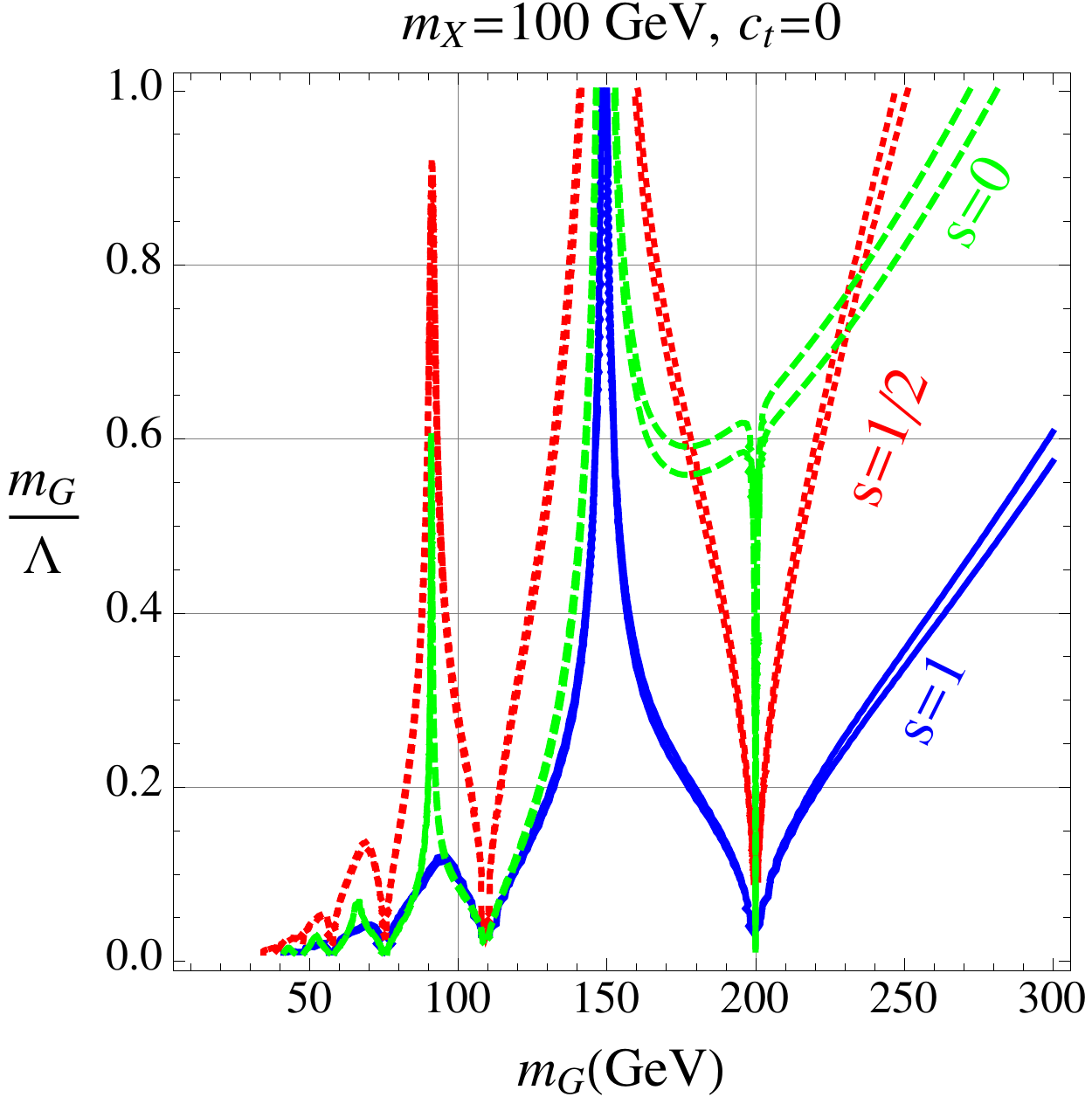}
\includegraphics[width=6cm]{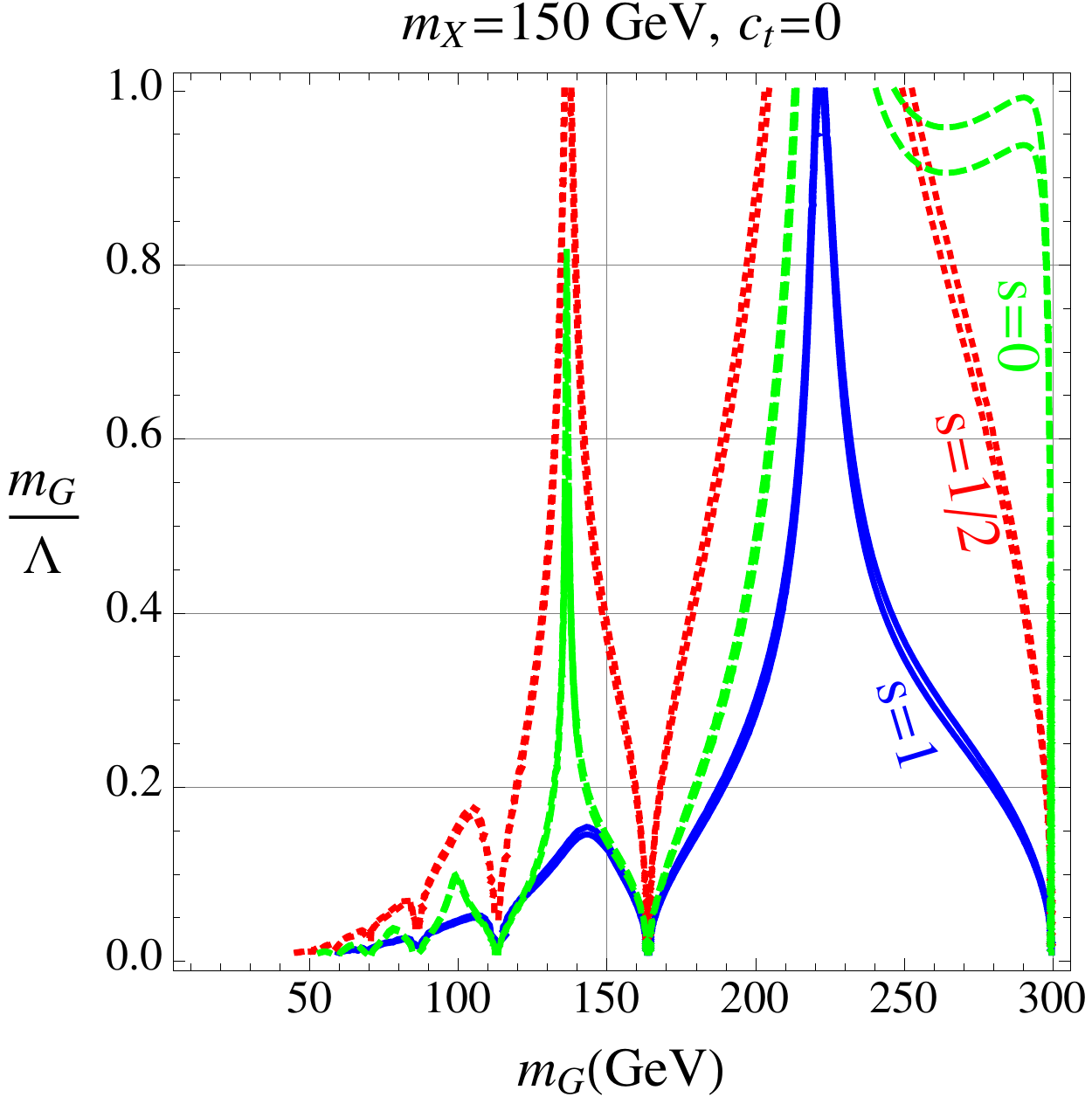}
\\ [5mm]
\includegraphics[width=6cm]{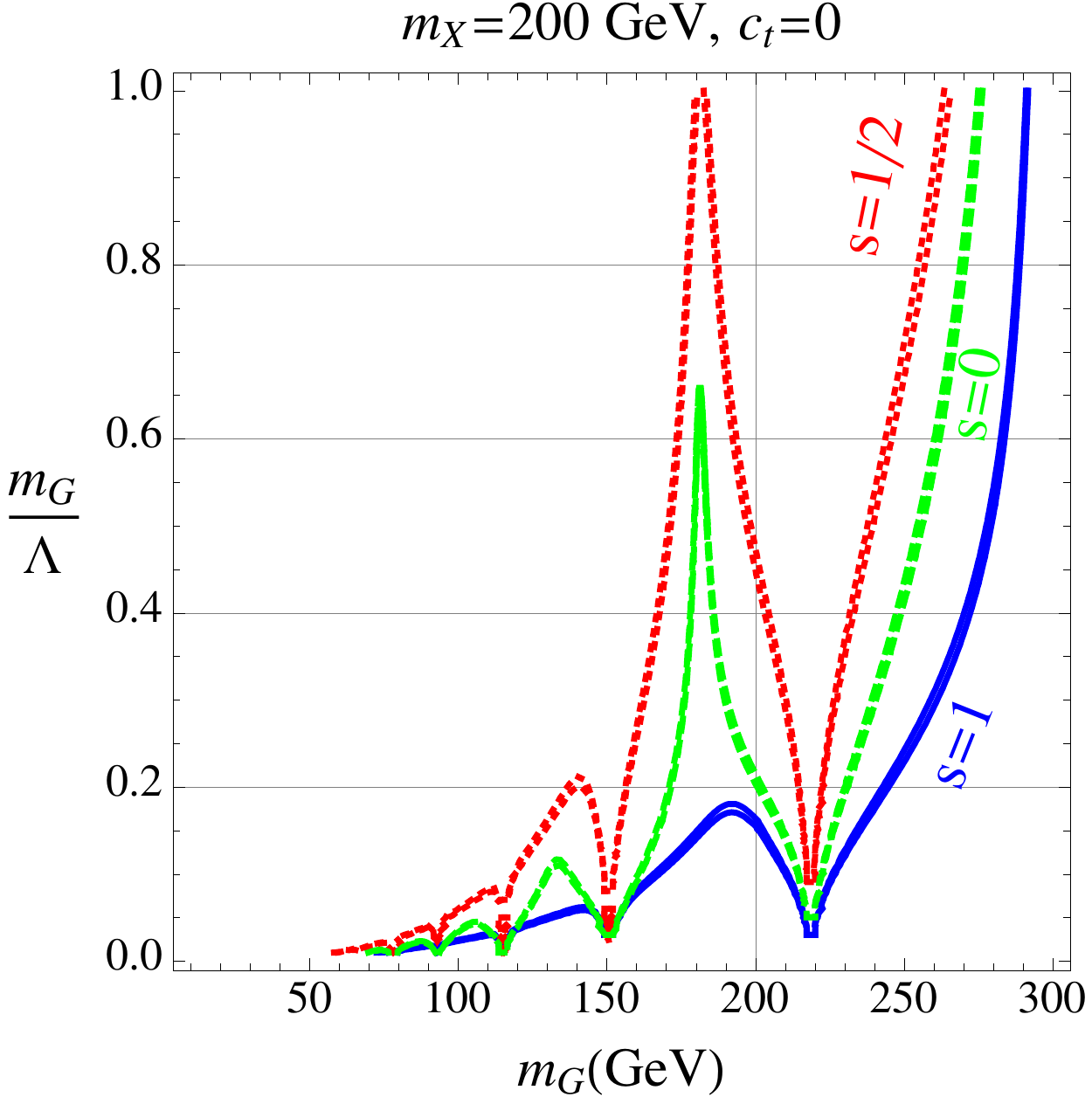}
\caption{Parameter space satisfying the relic density condition on the effective DM coupling, $m_G/\Lambda$, vs $m_G$, with or without $t{\bar t}$ channel.
We have set $m_X$=100, 150, 200 GeV as denoted in the plots. Blue (solid), red (dotted) and green (dashed) lines denote the Planck $5\sigma$ band for vector, fermion and scalar dark matters, respectively.
$c_X=1$, $c_H=1$, $c_V=0.03$.
}
\label{fig:relic1}
\end{figure}

\begin{figure}[t]
\centering
\includegraphics[width=6cm]{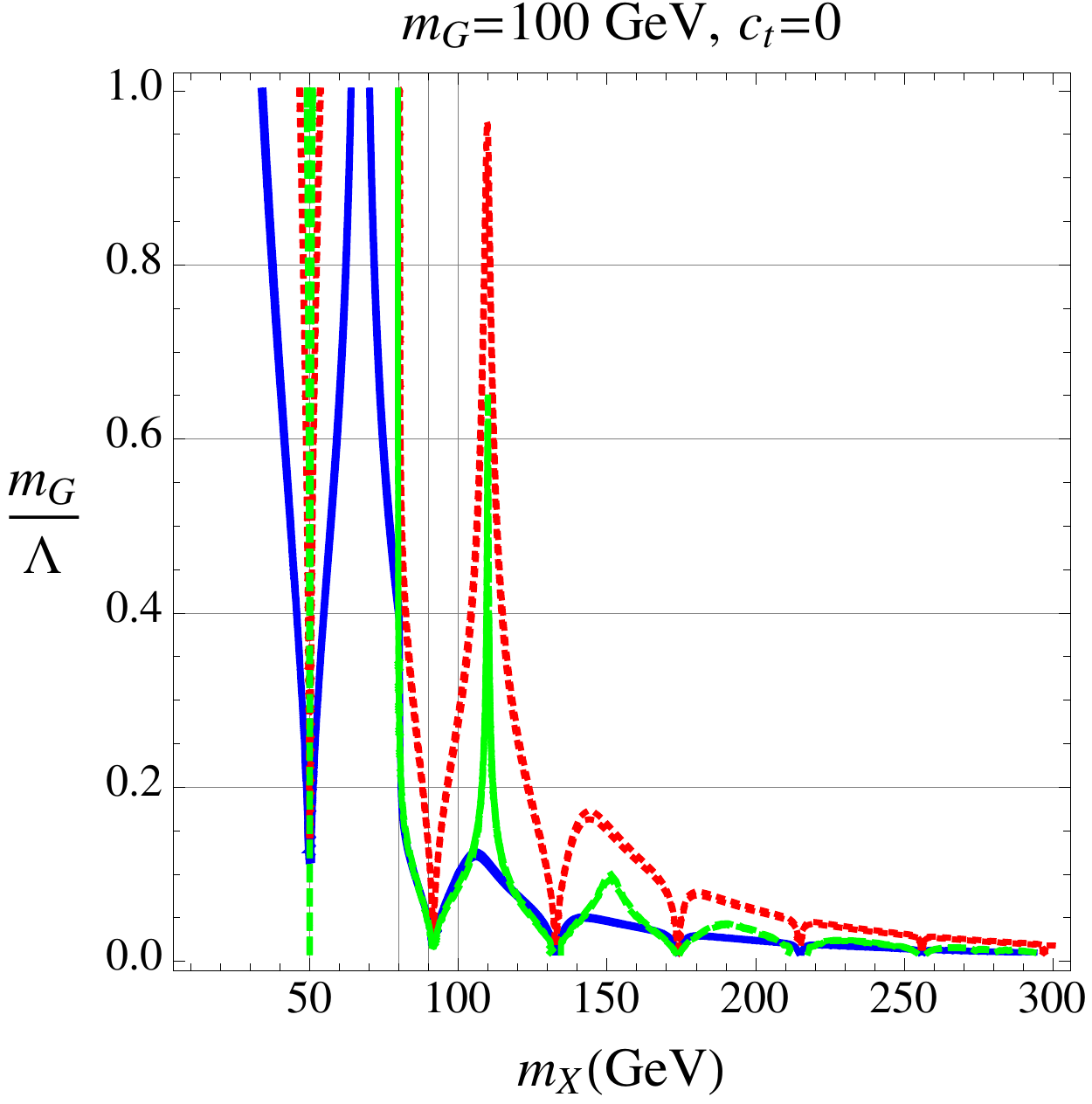}
\includegraphics[width=6cm]{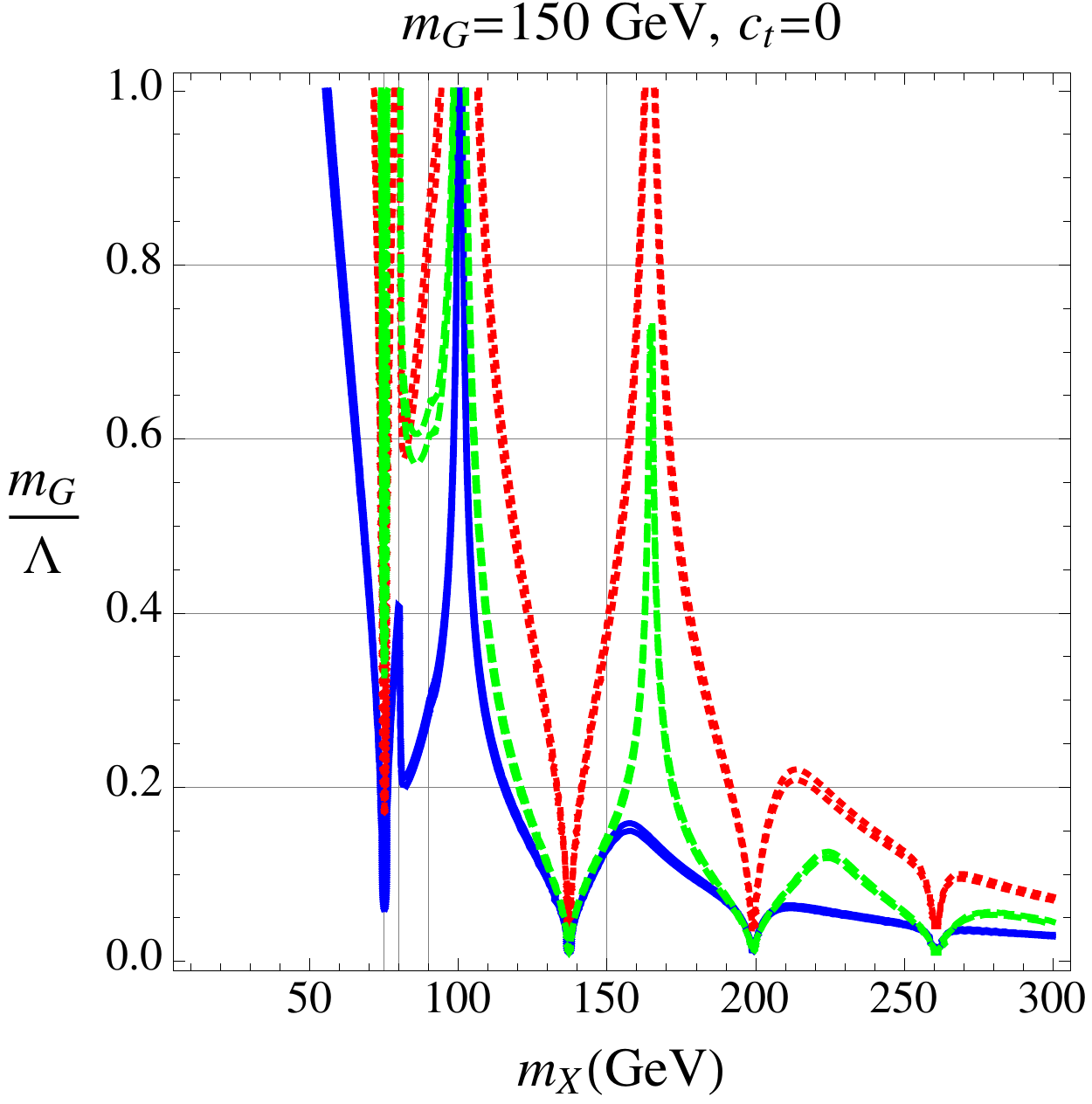}
\\ [5mm]
\includegraphics[width=6cm]{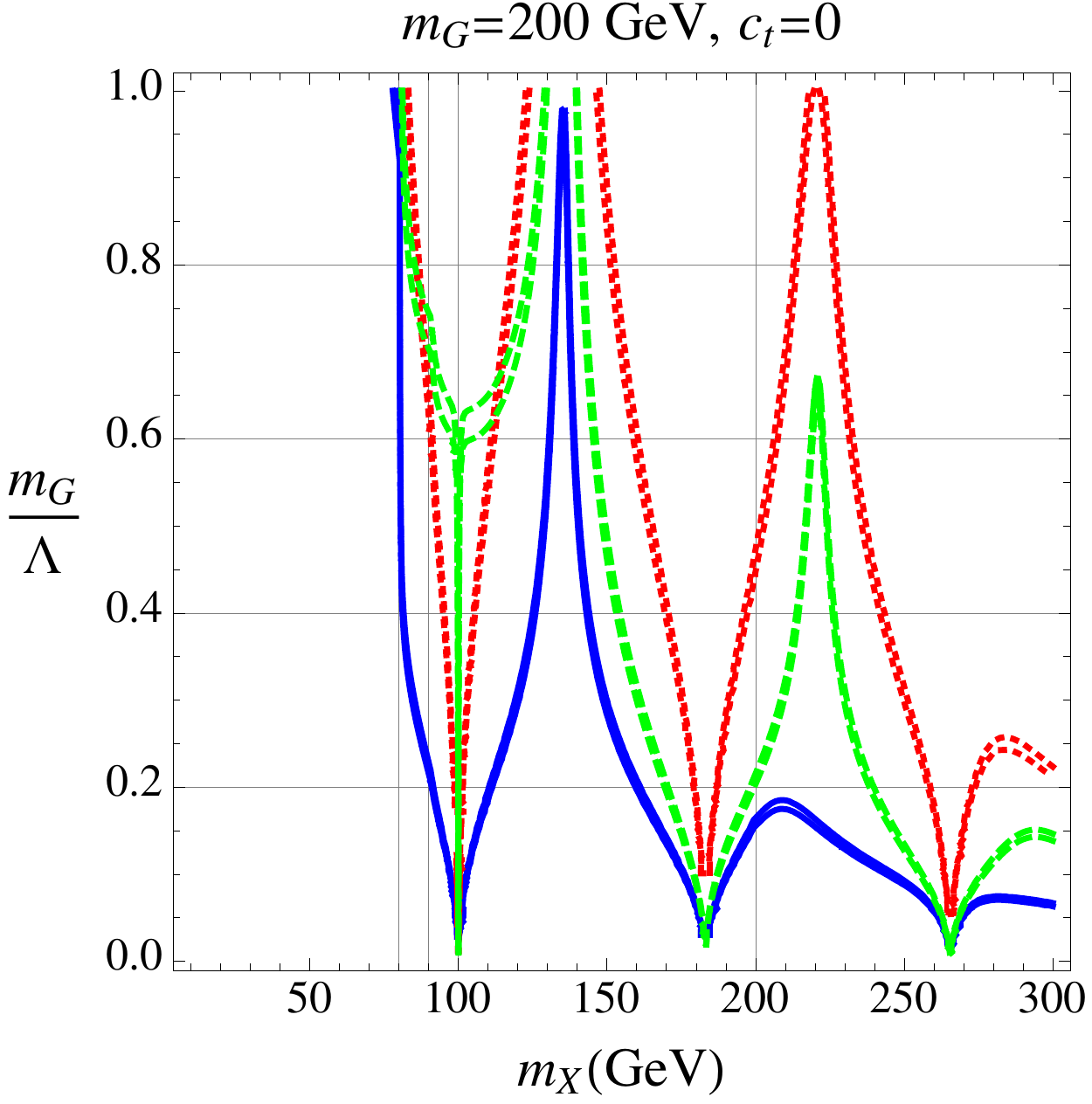}
\includegraphics[width=6cm]{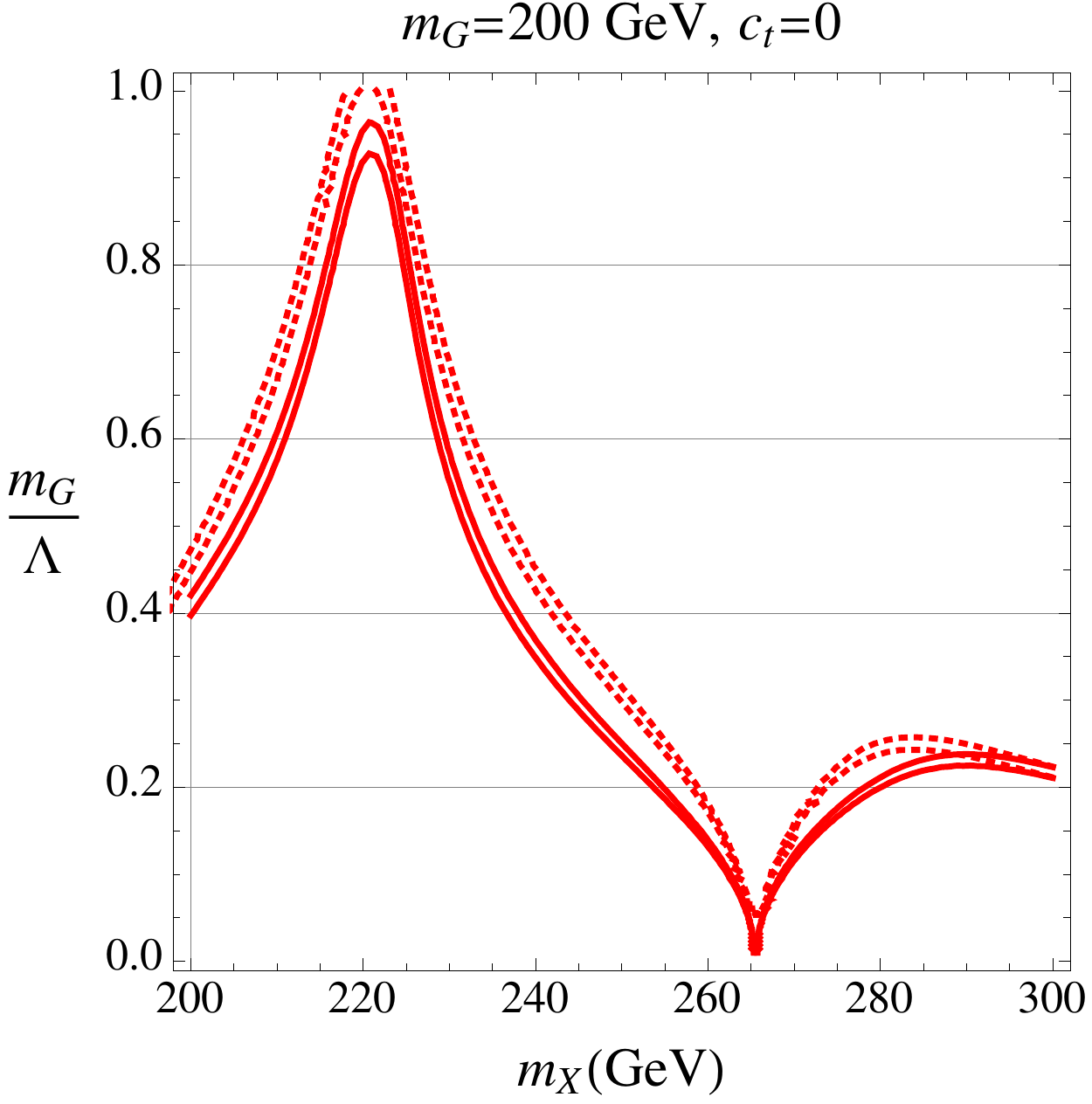}
\caption{Parameter space satisfying the relic density condition on the effective DM coupling, $m_G/\Lambda$, vs $m_X$, with or without  $t{\bar t}$ channel.
We have set $m_G=100, 150, 200\,{\rm GeV}$ as denoted in the plots. Blue (solid), red (dotted) and green (dashed) lines denote the Planck $5\sigma$ band for vector, fermion and scalar dark matters, respectively.
$c_X=1$, $c_H=1$, $c_V=0.03$ and $c_f=0$ are taken in common, except that $c_t=1$ in the right plot of the lower panel. In the right lower panel we show the effect of changing $c_t$ from 0 to 1. The effect is only sizeable in the large $m_X$ region.
}
\label{fig:relic2}
\end{figure}

In Figs.~\ref{fig:relic1} and \ref{fig:relic2}, we depict the parameter space on $m_G/\Lambda$ vs $m_G$ and $m_G/\Lambda$ vs $m_X$, respectively, for dark matter of $ s=0,1/2,1$ with KK graviton mediators, by considering the relic density condition obtained from Planck, 
$\Omega_{\rm DM}h^2=0.1199\pm 0.0027$ \cite{planck}. We have included the effect of the whole tower of KK gravitons in the AdS metric. The deeps in the plots correspond to the threshold of a new resonance and the spikes are due to the destructive interference between KK gravitons. For $m_G<2m_X$, the resonances occur at the higher KK modes too; for $m_G> 2m_X$, the higher KK modes are considered to be essentially decoupled for the DM annihilations, so it is a good approximation to take only the first KK graviton. There is a detailed discussion on the sum of KK gravitons in Appendix G.

We find that in all the dark matter cases, the relic density condition is satisfied for a wide range of the parameters. Note that in AdS models,  the dark matter coupling $m_G/\Lambda=3.8 k/M_{Pl}$.

In particular, for the case of vector dark matter,  a relatively small  effective DM coupling, $m_G/\Lambda$, is allowed due to the s-wave behavior of the annihilation cross section.  For instance, for $m_G$=100, 150, 200 GeV, the effective DM coupling can be smaller than $0.1$ away from the resonances, for $m_X\gtrsim$ 110, 180, 250 GeV, respectively. On the other hand, in the case of scalar dark matter, the annihilation cross section is suppressed by $m^4_{W,Z}/m^4_G$ for $m_G>m_{W,Z}$, as compared to the vector dark matter case, so it requires a larger effective dark matter coupling. 

Finally, for fermion dark matter, the annihilation cross section is p-wave suppressed, so it requires a larger effective DM gauge coupling. 
For instance, for $m_G=100(150)\,{\rm GeV}$, the effective DM coupling can be smaller than $0.1$, for $m_X\gtrsim 170(280)\,{\rm GeV}$.

The suppression of the DM annihilation cross sections is summarized in Table I, depending on the spin of dark matter and the type of mediator.

\subsection{Radion mediator}

In this paper we focus on the massive graviton as the mass is directly related to the compactification scale. The radion mass is a more model-dependent parameter, as it strongly depends on the stabilization mechanism, and it could be much heavier than the massive graviton. In this section we sketch, but do not provide details on the radion mediation. The computation of the annihilation cross section $X \bar{X} \to H H^{\dagger}$ follows the same steps as in the KK-graviton case, but with a simpler Lorentz structure. The result follows the general expression in Eq.~\ref{xstf}, with the substitution $m_G \to m_r$ and $c^G\to c^r$. Specifically,
\beq
(\sigma\vrel)_{X \bar{X} \to \phi \phi }   \sim \frac{(c_H^r c_{X}^r)^2}{\beta_s\pi\Lambda^4}\frac{ (a_s+b_s v^2+c_s v^4) m_{X}^6}{(m_r^2-4 m_{X}^2)^2+\Gamma_r^2 m_r^2} 
\eeq
where we have neglected terms of order ${\cal O}$($m_{\phi}^2/m_X^2$) and $\beta_s$ is a numerical constant which depends on the spin of $X$, see Appendix G.

Besides the  $X \bar{X} \to H H^{\dagger}$ processes, one could also consider $X \bar{X} \to r r$, a computation which was carried by the authors of Ref.~\cite{Bai-radion} in the limit $m_r \ll m_X$, and we refer the reader to this paper for details. Note, though, that Ref.~\cite{Bai-radion} uses a different parametrization of the radion quadratic couplings.

Whether the relic abundance is dominated by $X \bar{X} \to \phi \phi$ or $X \bar{X} \to r r $, one obtains the same velocity suppression, as shown in Table~\ref{table:collbounds}.

\subsection{Direct detection}

The interactions relevant to direct detection of dark matter are operators involving the first generation quarks. The coupling of $X$ to light fermions is very suppressed in this model, as the Dark Matter is fully composite and light fermions are elementary, and all the communication between them must go through a bulk field (partially composite). Since Dark Matter is a singlet of the SM, then the coupling is generated through gravity mediators. Indeed, the coupling to the SM quarks are suppressed by the exchange of KK-gravitons or radion, 
\bea
g_{X q} \propto \frac{c_{\psi} c_{X}}{m_G^2 \Lambda^2} \ ,
\eea 
hence very small and leading to no constraints from direct detection in the region of the parameter space consistent with the relic aboundance.

\begin{figure}[t!]
\centering
\includegraphics[width=6cm]{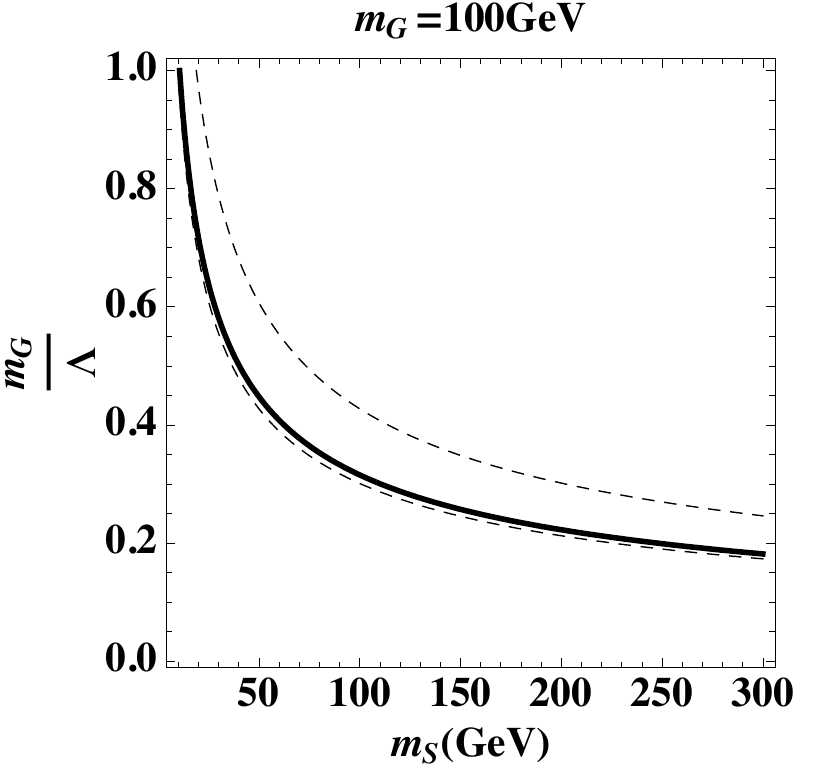}
\includegraphics[width=6cm]{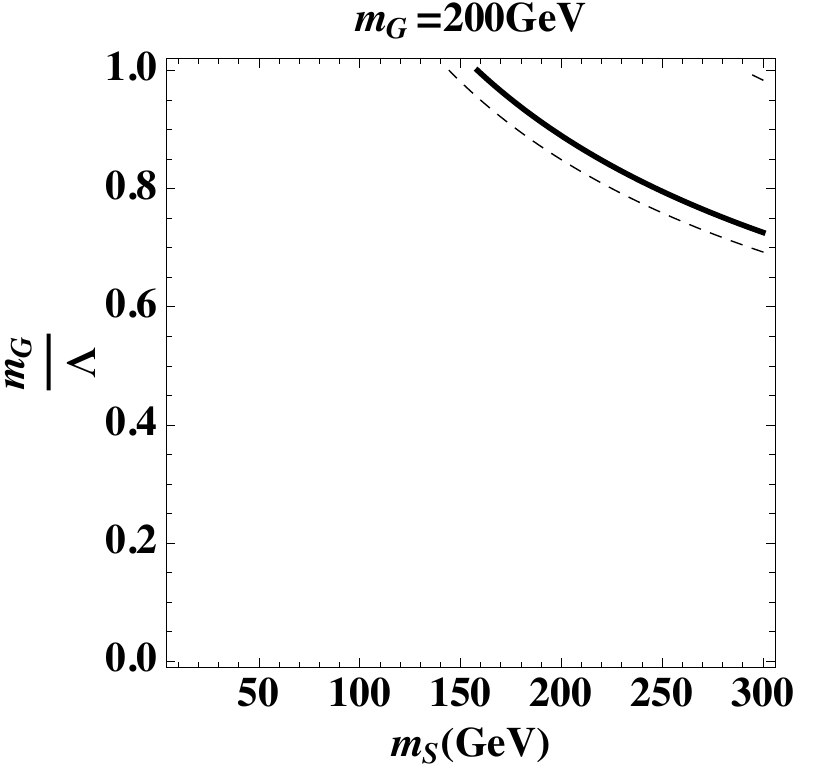}
\caption{Contours of spin-independent cross section of scalar dark matter on the parameter space, $m_S$ vs $m_G/\Lambda$. Solid line and dashed lines correspond to $\sigma_{S-N}=10^{-8}\,{\rm pb}$ for lattice and MILC results, respectively.
}
\label{directd}
\end{figure}

The effective four-point interactions between dark matter and Higgs, for instance, $S^2 h^2$, for scalar dark matter, are the strongest. But, their contributions to the spin-independent cross section are loop-suppressed, because there is no linear Higgs coupling to dark matter unlike Higgs portal couplings. The reason is that the KK graviton couples to the full Higgs potential through the energy-momentum tensor, where the linear term for the Higgs field vanishes due to the minimization condition. 
We also note that the interactions of dark matter to gluons could be most relevant for direct detection because gluons interact with dark matter more strongly than quarks. 
For instance, keeping the first KK graviton, the effective interactions between scalar dark matter and gluons are given by
\bea
{\cal L}_{S-N}=\xi_g \, S^2 G_{\mu\nu} G^{\mu\nu}.
\eea
with $\xi_g\equiv \frac{c_g c_S}{ 6\Lambda^2}\,\frac{m^2_S}{m^2_G}$. Then, the spin-independent cross section induced by the gluon interactions is 
\bea
\sigma_{S-N}= \frac{\mu^2}{\pi m^2_S} \left(\frac{8\pi}{9\alpha_S}\right)^2 m^2_N \xi^2_g f^2_{TG}
\eea
where $\mu=m_S m_N/(m_S+m_N)$ is the reduced mass of the nucleon-dark matter system and 
\bea
f_{TG}= \frac{1}{m_N} \langle N| \frac{-9\alpha_S}{8\pi} G_{\mu\nu} G^{\mu\nu}| N\rangle.
\eea
The lattice result gives $f_{TG}=0.867$ \cite{lattice} while the MILC results ranges between $0.472$ and $0.952$ \cite{MILC}.   As illustration, in Fig.~\ref{directd}, we depict the contours of the parameter space for scalar dark matter, giving rise to the spin-independent cross section, $\sigma_{S-N}=10^{-8}\,{\rm pb}$, depending on the results of the nucleon mass matrix. Consequently, direct detection bounds from XENON100 \cite{xenon100} can be strong enough to rule out a certain parameter space with light KK graviton and dark matter.

\section{Pushing the top to the bulk}\label{pusht}

The top quark could  directly participate in EWSB, and to what degree depends on the localization of the top in the extra-dimension. On the same token, the decay of the graviton or radion to tops depends on how localized the top is towards the Dark-brane. The localization is controlled by the bulk mass parameter~\cite{RSbulk}
\bea
{\cal L}_{5D} \supset M_{\Psi} \bar{\Psi} \Psi
\eea
where $\Psi$ is a 5D fermion.

It is convenient to define a dimensionless mass parameter $\nu_{\psi}=M/k$, where $k$ is the curvature of spacetime. As $\nu_{\psi}$ increases, the zero model is pushed toward the Dark-brane. For example, in AdS the effect of $\nu_{\psi}$ is as follows. For $\nu_{\psi}=1/2$ the fermion zero mode is de-localized in the extra dimension (flat profile), a point which is called the {\it conformal value}.  On the other hand, when $\nu_{\psi}>1/2$, the fermion zero-mode will be localized towards the  Dark-brane, whereas for  $\nu_{\psi}<1/2$, the localization is near the Matter-brane~\footnote{See Ref.~\cite{contino-ferm} for a discussion on the dual picture of fermion localization and the interpretation of the bulk mass $\nu_{\psi}$ from the point of view of partial compositeness. Note that fermion compositeness is not restricted to the third generation, and composite first and second generation fermions are possible from the point of view of flavour~\cite{andi} and leads to very distinctive signatures~\cite{compositejets}.}.

\begin{figure}[t!]
\begin{minipage}{8in}
\hspace*{-0.7in}
\centerline{\includegraphics[height=7cm]{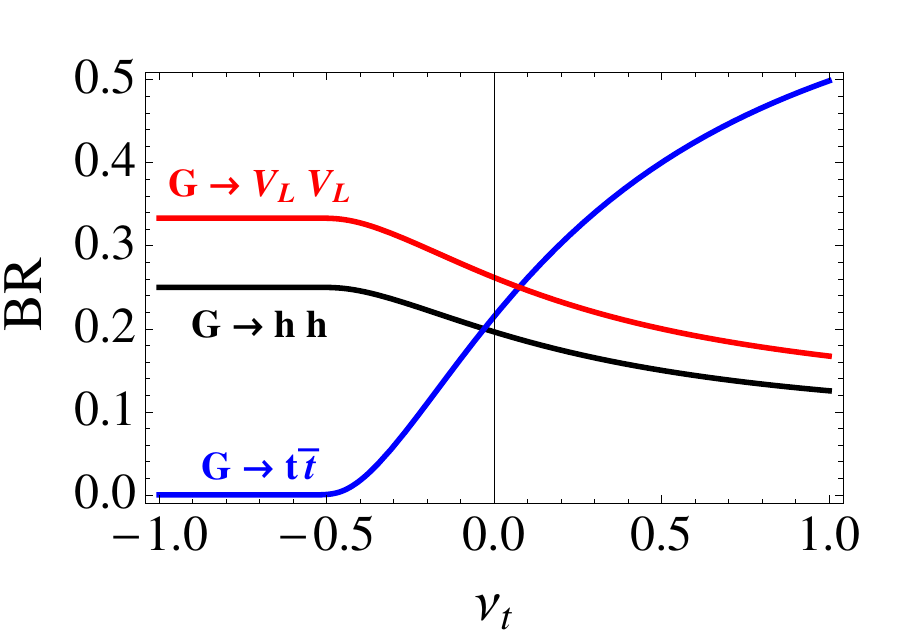}} \label{BRfig}
\hfill
\end{minipage}
\caption{
{\it
Branching ratio of graviton to the Higgs, vector bosons and top, as a function of the top bulk mass term, $\nu_t=M_{t}/k$. 
} 
}
\label{setup}
\end{figure}

As a result, the graviton and radion branching ratio (BR) to tops would depend on $\nu_{\psi}$: the larger the value of $\nu_{\psi}$, the larger the BR. For example, in AdS metrics the graviton BR to right-handed tops is given by 
\bea
\Gamma (G\to t \bar{t})  = \frac{ f(\nu_t)^2}{240 \, \pi}  \, m_G \left(\frac{m_G}{\Lambda} \right)^2
\eea
where we have neglected effects ${\cal O}(4 m_t^2/m_G^2)$, and $f(\nu_t)$ is defined as
\bea
f(\nu_t) =\frac{3}{2} \,  \frac{1+2 \nu_t}{1-e^{-k L (1+2 \nu_t)}} \, \int_0^1 d y y^{2+2 \nu_t} \frac{J_2(3.83 y)}{J_2(3.83)} \ .
\eea

In Fig.~\ref{BRfig}, we show the BR of graviton to the Higgs degrees of freedom ($h$, $Z_L$ and $W^{\pm}_L$) and to tops. If the top is pushed towards the Dark-brane, the graviton decay to tops could dominate, but depends crucially on the degree of localization.

When the top quark is localized on the Dark brane as well, dark matter can annihilate sizably into a top quark pair if kinematically allowed.
From the results in Appendx B, depending on the spin of dark matter, the annihilation cross sections are given by 
\bea
(\sigma v)_{SS\rightarrow t{\bar t}}&\simeq&\frac{(c^G_S c^G_t)^2v^4}{1920\pi\Lambda^4}\, m^2_S, \\
(\sigma v)_{\chi{\bar\chi}\rightarrow t{\bar t}}&\simeq&\frac{(c^G_\chi c^G_t)^2v^2}{384\pi\Lambda^4}\, m^2_\chi, \\
(\sigma v)_{XX\rightarrow t{\bar t}}&\simeq&\frac{(c^G_X c^G_t)^2}{36\pi\Lambda^4}\, m^2_X \ ,
\eea
in the limit of $m_X\gg m_t, m_G$.
Thus, the $t{\bar t}$ channel can give a sizable contribution to the annihilation cross sections of fermion and vector dark matters, while it becomes d-wave and negligible for scalar dark matter. Consequently,  in the case of fermion and vector dark matters, the relic density condition needs a smaller effective dark matter coupling to the KK graviton, $m_G/\Lambda$, due to the localization of the top quark on the Dark brane.
The effect of the $t{\bar t}$ channel on the relic density is shown in the right plot of the lower panel In Figs.~\ref{fig:relic1} and \ref{fig:relic2}.

\section{Collider searches}\label{collider}

\begin{figure}[t!]
\hspace*{-.70in}
\begin{minipage}{8in}
\hspace*{-0.7in}
\includegraphics[height=5cm]{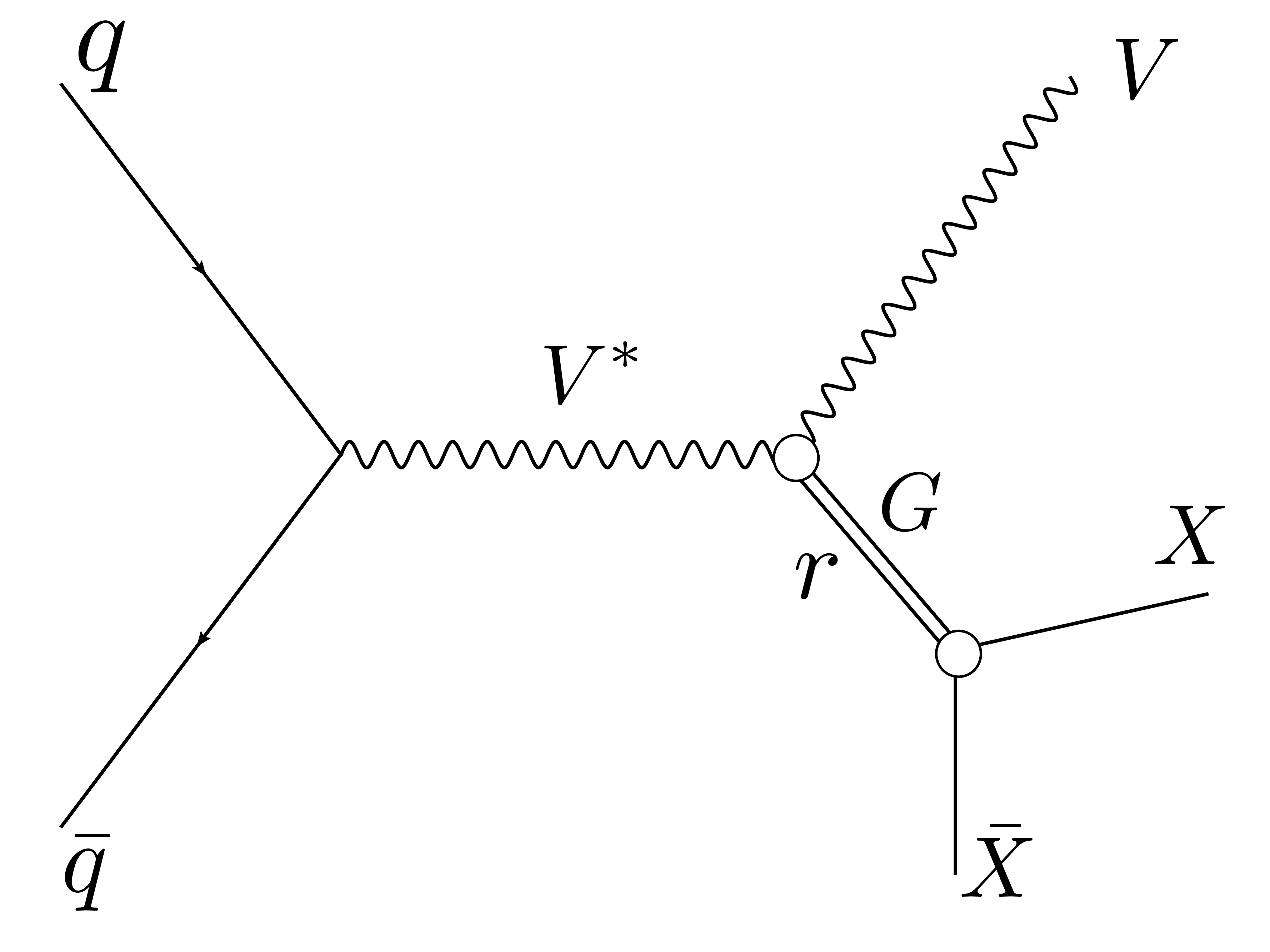}
\includegraphics[height=5.5cm]{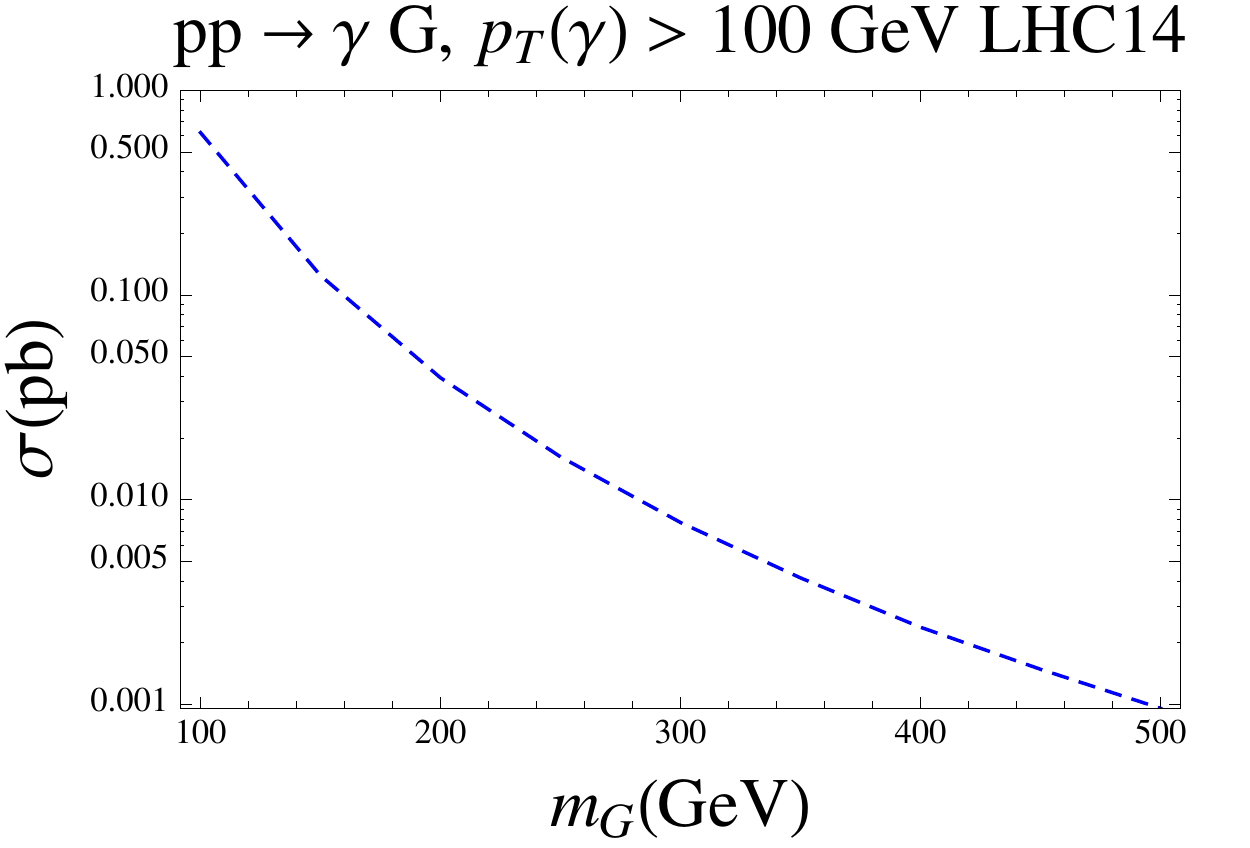}
 \label{assocfig}
\hfill
\end{minipage}
\caption{
{\it
Production of gravity mediators in association with a vector boson at the LHC.  
} 
}
\label{setup}
\end{figure}

Radion and KK-gravitons searches at the LHC are based on assumptions about the decay of those particles, which does not match this model. For example, bounds on the radion mass  compiled in Refs.~\cite{barger,sr} assume an amount of mixing between the Higgs and the radion. Similarly, experimental searches of extra-dimensions~\cite{CMSGG,VVboosted, CMSdilep, ATLASdilep, boostedtops} are based on decays to photons, leptons and four-fermion operators involving light fermions, which are very suppressed in this scenario, or very boosted tops, which are only valid for $m_G\gtrsim 1$ TeV. Other indirect constraints, such as loop contributions to precision electroweak parameters, are also very mild, see Ref.~\cite{Tao} for a study in the case of Universal Extra-Dimensions, i.e. flat extra-dimensions.

Instead, the best channels to look for the radion and graviton in this model is via  production of the mediator in association with a gauge boson, see Fig.~\ref{assocfig}.  The signatures would be missing energy with mono-Z~\cite{LHCmonoZ,mono-Z}, mono-lepton~\cite{mono-lepton} and mono-photon~\cite{monoa}. Vector boson fusion~\cite{vbf} would be suppressed at low dark matter mediator mass respect to associated production, but a promising channel at high mass.

Searches for mediators in monophoton events~\cite{CMS-monoa,ATLAS-monoa} can be re-interpreted in terms of the process in Fig.~\ref{assocfig}. LHC at 14 TeV might be sensitive to the coupling of gravity mediators to photons (see Eq.~\ref{cgamma}).  For illustration purposes, we show in Fig.~\ref{assocfig} (right) the production cross section (in pb's) of a graviton in association with a photon, with a cut on photon $p_T$ of 100 GeV at LHC14. The numbers correspond to a choice of $\Lambda$= 1 TeV, and re-scaling to other values of $\Lambda$ is trivial. As this study is beyond the scope of this paper, we leave it for a future publication.

\section{Conclusion}

Gravity could communicate the Dark Matter sector with the visible (SM) sector via gravity mediators. Those mediators (KK-gravitons, radions) are a consequence of extra-dimensions which are compactified. In this paper we show that such scenarios, compatible with a solution to the hierarchy problem, can comfortably accommodate the observed relic abundance and yet be safe from direct detection constraints.

Most interestingly, this scenario is not exclusive of extra-dimensional models. Despite the name, Gravity-mediated Dark Matter is also a mechanism which could arise from a strongly coupled, near-conformal scenario. We have developed this dual picture, based on the breaking of conformality and partial compositeness, obtaining that the computation of dark matter relic abundance in the gravitational side can be exactly matched to the holographic four-dimensional model.

\section*{Acknowledgments}
VS thanks Carlos Nu\~nez for enlightening discussions.
The work of VS is supported by the Science Technology and Facilities Council (STFC) under grant number ST/J000477/1. The work of HML is supported in part by Basic Science Research Program through the National Research Foundation of Korea(NRF) funded by the Ministry of Education, Science and Technology(2013R1A1A2007919). The work of MP is supported by a CERN-Korean fellowship.

\def\theequation{A.\arabic{equation}}

\setcounter{equation}{0}

\vskip0.8cm
\noindent
{\Large \bf Appendix A: Spin-2 massive graviton} 
\vskip0.4cm
\noindent

In unitary gauge~\cite{giudice,lykken,Hagiwara}, the propagator of spin-2 massive graviton with momentum $k$ from $G_{\mu\nu}$ to $G_{\alpha\beta}$ is 
\beq
i \Delta^G_{\mu\nu,\alpha\beta}(k) = \frac{i P_{\mu\nu ,\alpha\beta}(k)}{k^2-m^2} ,
\eeq 
and the spin-sum of the polarization tensors is
\be
\sum_s \epsilon_{\mu\nu}(k,s) \epsilon_{\alpha\beta}(k,s)=P_{\mu\nu,\alpha\beta}(k),
\ee
where 
\be
P_{\mu\nu,\alpha\beta}(k)=\frac{1}{2}\Big(G_{\mu\alpha}G_{\nu\beta}+G_{\nu\alpha}G_{\mu\beta}-\frac{2}{3}G_{\mu\nu} G_{\alpha\beta}\Big) \label{PropG}
\ee
with
\be
G_{\mu\nu}\equiv \eta_{\mu\nu}-\frac{k_\mu k_\nu}{m^2_G}.
\ee
The tensor $P_{\mu\nu\alpha\beta}$ satisfies  traceless and transverse conditions for an on-shell graviton $G_{\mu\nu}$ case as following, 
\bea
\eta^{\alpha\beta} P_{\mu\nu,\alpha\beta}(k)&=&0, \\
k^\alpha P_{\mu\nu,\alpha\beta}(k)&=&0.
\eea

The energy-momentum tensor for the SM and dark matter is given by

\bea
T_{\mu\nu}&=&T^{\rm SM}_{\mu\nu}+T^{\rm DM}_{\mu\nu}
\eea
with
\bea
T^{\rm SM}_{\mu\nu}&=& \left[\frac{i}{4}{\bar\psi}(\gamma_\mu D_\nu+\gamma_\nu D_\mu)\psi-\frac{i}{4}(D_\mu{\bar\psi}\gamma_\nu+D_\nu{\bar\psi}\gamma_\nu)\psi 
-g_{\mu\nu} ({\bar\psi}\gamma^\mu D_\mu\psi-m_\psi {\bar\psi}\psi)+\frac{i}{2}g_{\mu\nu}\partial^\rho({\bar\psi}\gamma_\rho\psi ) \right]\nonumber \\
&+&\left[\frac{1}{4}g_{\mu\nu} F^{\lambda\rho}F_{\lambda\rho}-F_{\mu\lambda}F^\lambda\,_\nu\right]
+\left[-g_{\mu\nu}D^\rho H^\dagger D_\rho H+g_{\mu\nu}V(H)+D_\mu H^\dagger D_\nu H+D_\nu H^\dagger D_\mu H\right], ~~~~
 \label{fullTmunu}
\eea 
\bea
T^{\rm (Vector~DM)}_{\mu\nu}&=&
 \frac{1}{4}g_{\mu\nu} X^{\lambda\rho}X_{\lambda\rho}+X_{\mu\lambda}X^\lambda\,_\nu+m^2_X\Big(X_\mu X_\nu-\frac{1}{2}g_{\mu\nu} X^\lambda  X_\lambda\Big)  , \nonumber\\
T^{\rm (Fermion~DM)}_{\mu\nu}&=& \frac{i}{4}{\bar\chi}(\gamma_\mu\partial_\nu+\gamma_\nu\partial_\mu)\chi-\frac{i}{4} (\partial_\mu{\bar\chi}\gamma_\nu+\partial_\nu{\bar\chi}\gamma_\nu)\chi-g_{\mu\nu}(i {\bar\chi}\gamma^\mu\partial_\mu\chi- m_\chi {\bar\chi}\chi) 
+\frac{i}{2}g_{\mu\nu}\partial^\rho({\bar\chi}\gamma_\rho\chi),  \nonumber \\
T^{\rm (Scalar~DM)}_{\mu\nu}&=& \partial_\mu S \partial_\nu S-\frac{1}{2}g_{\mu\nu}\partial^\rho S \partial_\rho S+\frac{1}{2}g_{\mu\nu}  m^2_S S^2 , 
\eea
KK graviton couples with SM and DM particles through energy momentum tensors with $\frac{c_i}{\Lambda}$ couplings.
Here, $\Lambda$ is the cutoff scale which is taken to be larger than the KK graviton mass.
$c_i$ will be one of $\{c_{X,S,\chi}, c_A, c_\psi, c_H\}$ depending on a particle that are determined by the overlaps between the wave functions of KK graviton and matter fields in extra dimensions \cite{RSbulk}. 

\def\theequation{B.\arabic{equation}}

\setcounter{equation}{0}

\vskip0.8cm
\noindent
{\Large \bf Appendix B: Decay rates of KK graviton} 
\vskip0.4cm
\noindent

In this appendix, we present the details of the KK graviton decay rates. We follow the conventions for the KK graviton propagator and interactions in Ref.~\cite{lykken}.
The vertex Feynman rules between incoming KK graviton and outgoing scalar particles with momentum $k_1$ and $k_2$ will be
\bea
&&\left[G_{\mu\nu}, S(k_1) , S(k_2) \right]: -i\frac{c_S}{\Lambda}\left(m^2_S \eta_{\mu\nu}-C_{\mu\nu,\rho\sigma}k_1^\rho k_2^\sigma\right) \\
&&\left[G_{\mu\nu}, h(k_1) , h(k_2) \right]: -i\frac{c_H}{\Lambda} \left(m^2_h \eta_{\mu\nu}-C_{\mu\nu,\rho\sigma}k_1^\rho k_2^\sigma\right).
\eea
Similarly for the incoming graviton and outgoing massive vector bosons case is following,
\bea
&&\left[G_{\mu\nu}, V_\alpha(k_1) , V_\beta(k_2) \right]:-i \frac{1}{\Lambda} ( c_H m^2_A C_{\mu\nu,\alpha\beta}+c_V W_{\mu\nu,\alpha\beta}) 
\textrm{: Gauge boson case}\\
&&\left[G_{\mu\nu}, X_\alpha(k_1) , X_\beta(k_2) \right]:-i \frac{c_V}{\Lambda} ( m^2_A C_{\mu\nu,\alpha\beta}+ W_{\mu\nu,\alpha\beta})
\textrm{: Dark matter case}
\eea
depending on a mass mechanism for a vector boson. When a mass term for a vector boson is generated by higgs mechanism like standard model vector bosons, 
a graviton couples a mass term with a different coupling constant $c_H$ compared to gauge kinematic terms.
with 
\bea
W_{\mu\nu,\alpha\beta} &\equiv&\eta_{\alpha\beta}k_{1\mu}k_{2\nu}+\eta_{\mu\alpha}(k_1\cdot k_2\,\eta_{\nu\beta}-k_{1\beta}k_{2\nu} )-\eta_{\mu\beta}k_{1\nu}k_{2\alpha}+\frac{1}{2}\eta_{\mu\nu}(k_{1\beta}k_{2\alpha}-k_1\cdot k_2\, \eta_{\alpha\beta}  )+(\mu\leftrightarrow \nu) , \nonumber \\
C_{\mu\nu,\alpha\beta} &\equiv&\eta_{\mu\alpha}\eta_{\nu\beta}+\eta_{\nu\alpha}\eta_{\mu\beta}-\eta_{\mu\nu}\eta_{\alpha\beta} 
\eea
 respectively. Here, we took the unitary gauge for gauge bosons.
The Feynman rule between incoming graviton and outgoing fermions and anti fermion is 
\beq 
&&\left[G_{\mu\nu}, \bar \psi(k_1) , \psi(k_2) \right]: 
-i\frac{c_\psi}{4\Lambda}\left(\gamma_\mu (k_{1\nu}-k_{2\nu})+\gamma_\nu(k_{1\mu}-k_{2\mu})-2\eta_{\mu\nu}(\slashed k_1-\slashed k_2-2m_\psi )  \right).
\eeq

The decay amplitude for $G_{KK}\rightarrow h(k_1)h(k_2)$ is
 
\beq
\Gamma(G_{KK}\rightarrow hh)= \frac{c^2_H m^3_G}{960 \pi \Lambda^2} \Big(1-\frac{4m^2_h}{m^2_G}\Big)^\frac{5}{2}.\label{gammahh}
\eeq

Next,  the decay amplitude into massive gauge bosons, $G_{KK}\rightarrow A(k_1) A(k_2)$ is
\bea
\Gamma(G_{KK}\rightarrow ZZ)&=& \frac{m^3_G}{960 \pi \Lambda^2}
\bigg[c^2_H\Big(1+\frac{12m^2_Z}{m^2_G}+\frac{56m^4_Z}{m^4_G}\Big)+80 c_V c_H \Big(1-\frac{m^2_Z}{m^2_G}\Big)\frac{m^2_Z}{m^2_G} \nonumber \\
&&+12c^2_V\Big(1-\frac{3m^2_Z}{m^2_G}+\frac{6m^4_Z}{m^4_G}\Big)  \bigg]\Big(1-\frac{4m^2_Z}{m^2_G}\Big)^\frac{1}{2}, \\
\Gamma(G_{KK}\rightarrow WW)&=&\frac{m^3_G}{480 \pi \Lambda^2}
\bigg[c^2_H\Big(1+\frac{12m^2_W}{m^2_G}+\frac{56m^4_W}{m^4_G}\Big)+80 c_V c_H \Big(1-\frac{m^2_W}{m^2_G}\Big)\frac{m^2_W}{m^2_G} \nonumber \\
&&+12c^2_V\Big(1-\frac{3m^2_W}{m^2_G}+\frac{6m^4_W}{m^4_G}\Big)  \bigg]\Big(1-\frac{4m^2_W}{m^2_G}\Big)^\frac{1}{2}.
\eea

When $c_V=c_H$, from the above results, the decay rates into a pair of massive gauge bosons become
\bea
\Gamma(G_{KK}\rightarrow Z Z)&=& \frac{c^2_V m^3_G}{960\pi \Lambda^2}\Big(1- \frac{4m^2_Z}{m^2_G}\Big)^\frac{1}{2}
\Big(13+\frac{56m^2_Z}{m^2_G}+\frac{48m^4_Z}{m^4_G}\Big), \\
\Gamma(G_{KK}\rightarrow W W )&=& \frac{c^2_V m^3_G}{480\pi \Lambda^2}\Big(1- \frac{4m^2_W}{m^2_G}\Big)^\frac{1}{2}
\Big(13+ \frac{56m^2_W}{m^2_G}+\frac{48m^4_W}{m^4_G}\Big).
\eea
On the other hand, for $c_H=0$ and $m_A=0$, the decay rates for the KK graviton into a photon pair or a gluon pair follow
\bea
\Gamma(G_{KK}\rightarrow\gamma\gamma)&=& \frac{c^2_\gamma m^3_G}{80\pi\Lambda^2}, \\
\Gamma(G_{KK}\rightarrow gg)&=&  \frac{c^2_g m^3_G}{10\pi\Lambda^2}.
\eea

Lastly, the decay amplitude squared into a fermion pair, $G_{KK}\rightarrow \psi{\bar\psi}$ is
 \bea
 \Gamma(G_{KK}\rightarrow\psi  \bar \psi)   = \frac{c_\psi^2 m_G}{160 \pi}
 \left(\frac{m_G}{\Lambda}\right)^2 
 \left(1-\frac{4m^2_\psi}{m^2_G} \right)^\frac{3}{2} \left(1+\frac{8}{3} \frac{m^2_\psi}{m^2_G}\right) .
 \eea

\vskip0.8cm
\noindent
{\Large \bf Appendix C:  DM annihilation cross sections in scalar dark matter case} 
\vskip0.4cm
\noindent

In this section, we present the results of the annihilation cross sections for scalar dark matter.
Using the non-relativistic limit where $v_{DM}\ll 1$,  where
\bea
&& s \simeq m_S^2(4+v_\textrm{rel}^2) ,\\
&&k_1\cdot k_4 = k_2\cdot k_3 \simeq m_S^2 +\frac{1}{2} m_S^2\sqrt{1-\frac{m_h^2}{m_S^2}} 
\cos{\theta}\cdot \vrel +\frac{1}{4} m_S^2 \cdot \vrel^2, \\
&&k_1\cdot k_3 = k_2\cdot k_4 \simeq m_S^2 -\frac{1}{2} m_S^2\sqrt{1-\frac{m_h^2}{m_S^2}} \cos{\theta}\cdot \vrel 
+\frac{1}{4} m_S^2\cdot \vrel^2,
\eea
with an angle $\theta$ of $k_3$ with a respect to a direction of $k_1$ at the CM frame of $SS$ collision.
In this limit, we first consider DM annihilation cross sections for scalar dark matter case.
An annihilation cross section times relative velocity as
\beq
(\sigma \vrel)_{SS\rightarrow hh} \simeq \vrel^4 \cdot\frac{ c^2_S c^2_H}{720\pi \Lambda^4}\, \frac{m^6_S}{(4m^2_S-m^2_G)^2+\Gamma^2_G m^2_G} \left(1-\frac{m^2_h}{m^2_S}\right)^\frac{5}{2}.
\eeq

Similarly,  the amplitude for a scalar DM pair annihilating into a pair of massive gauge bosons is
\bea
(\sigma \vrel)_{SS\rightarrow Z Z}&\simeq& \frac{3c^2_S(c_V-c_H)^2}{16\pi \Lambda^4} \frac{m^2_S m^4_Z}{(4m^2_S-m^2_G)^2+\Gamma^2_G m^2_G}\left(1-\frac{4m^2_S}{m^2_G}\right)^2\left(1-\frac{m^2_Z}{m^2_S}\right)^{\frac{1}{2}},  \\
(\sigma \vrel)_{SS\rightarrow W W}&\simeq& \frac{3c^2_S(c_V-c_H)^2}{8\pi \Lambda^4} \frac{m^2_S m^4_W}{(4m^2_S-m^2_G)^2+\Gamma^2_G m^2_G}\left(1-\frac{4m^2_S}{m^2_G}\right)^2\left(1-\frac{m^2_Z}{m^2_S}\right)^{\frac{1}{2}}.
\eea
For $c_H=c_V$, both s-wave and p-wave components are zero and the annihilation cross section becomes d-wave as
\bea
(\sigma \vrel)_{SS\rightarrow Z Z}&\simeq&  \vrel^4 \cdot 
\frac{c^2_S c^2_V}{720\pi \Lambda^4} \frac{m^6_S}{(4m^2_S-m^2_G)^2+\Gamma^2_G m^2_G}\left(1-\frac{m^2_Z}{m^2_S}\right)^{\frac{1}{2}}
\left(13+\frac{14m^2_Z}{m^2_S}+\frac{3m^4_Z}{m^4_S}\right), \\
(\sigma \vrel)_{SS\rightarrow WW}&\simeq&  \vrel^4 \cdot
\frac{c^2_S c^2_V }{360\pi \Lambda^4} \frac{m^6_S}{(4m^2_S-m^2_G)^2+\Gamma^2_G m^2_G}\left(1-\frac{m^2_W}{m^2_S}\right)^\frac{1}{2}
\left(13+\frac{14m^2_W}{m^2_S}+\frac{3m^4_W}{m^4_S}\right). \quad
\eea
We also find that the annihilation cross sections into a photon pair or a gluon pair are always d-wave and are given by
\bea
(\sigma \vrel)_{SS\rightarrow \gamma\gamma}&\simeq&\vrel^4 \cdot  \frac{c^2_S c^2_\gamma }{60\pi\Lambda^4}\frac{m^6_S}{(4m^2_S-m^2_G)^2+\Gamma^2_G m^2_G},\\
(\sigma \vrel)_{SS\rightarrow gg}&\simeq& \vrel^4 \cdot  \frac{2c^2_S c^2_g }{15\pi\Lambda^4}\frac{m^6_S}{(4m^2_S-m^2_G)^2+\Gamma^2_G m^2_G}.
\eea

Finally,a dark matter annihilating into a pair of massive fermions case, an annihilation cross section is 
\bea
(\sigma \vrel)_{SS\rightarrow\psi{\bar\psi} } \simeq \vrel^4 \cdot  \frac{ c_S^2 c_\psi^2 }{360\pi \Lambda^4} 
\frac{m_S^6}{(m_G^2-4 m_S^2)^2+\Gamma_G^2 m_G^2}
\left(1-\frac{m_\psi^2}{m_S^2}\right)^\frac{3}{2} \left(3+\frac{2m_\psi^2}{m_S^2}\right) .
\eea
As in $S,S \rightarrow h,h$ case, there is a cancellation between on-shell and off-shell graviton contribution, so the $SS\rightarrow \psi{\bar\psi}$ is d-wave.

\def\theequation{D.\arabic{equation}}

\setcounter{equation}{0}

\vskip0.8cm
\noindent
{\Large \bf Appendix D:  DM annihilation cross sections in fermion dark matter case} 
\vskip0.4cm
\noindent

In this section, we present the results of the annihilation cross sections for fermion dark matter 
\bea
(\sigma \vrel)_{\chi{\bar\chi}\rightarrow hh}\simeq \vrel^2\cdot \frac{c^2_\chi c^2_H }{144\pi\Lambda^4}
\frac{m^6_\chi}{(4m^2_\chi-m^2_G)^2+\Gamma^2_G m^2_G} \left(1-\frac{m^2_h}{m^2_\chi}\right)^\frac{5}{2}.
\eea
Thus, the resulting annihilation cross section is p-wave.

The annihilation cross sections for fermion dark matter going into a pair of massive gauge bosons, $\chi{\bar\chi}\rightarrow AA$, are
\bea
(\sigma \vrel)_{\chi{\bar\chi}\rightarrow ZZ} &\simeq&  \vrel^2 \cdot \frac{c_\chi^2 c_V^2}{144\pi \Lambda^4} 
\frac{m_\chi^6}{(m_G^2-4 m_\chi^2)^2+\Gamma_G^2 m_G^2} \Bigg[\left(13+\frac{14m_Z^2}{m_\chi^2}+\frac{3m_Z^4}{m_\chi^4}\right)
-2  \left(1-\frac{c_H}{c_V}\right) \left(1+\frac{13m_Z^2}{m_\chi^2}+\frac{m_Z^4}{m_\chi^4}\right)\nonumber \\
&&+ \left(1-\frac{c_H}{c_V}\right)^2 \left\{1+\frac{3m_Z^2}{m_\chi^2}+\frac{31}{8}\frac{m_Z^4}{m_\chi^4}
-\frac{3m_Z^4}{m_G^2 m_\chi^2}+\frac{6m_Z^4}{m_G^4}
\right\}
\Bigg] \left(1-\frac{m_Z^2}{m_\chi^2}\right)^\frac{1}{2} , \\
(\sigma \vrel)_{\chi{\bar\chi}\rightarrow WW} &\simeq&  \vrel^2 \cdot \frac{c_\chi^2 c_V^2}{72\pi \Lambda^4} 
\frac{m_\chi^6}{(m_G^2-4 m_\chi^2)^2+\Gamma_G^2 m_G^2} \Bigg[\left(13+\frac{14m_W^2}{m_\chi^2}+\frac{3m_W^4}{m_\chi^4}\right)
-2  \left(1-\frac{c_H}{c_V}\right) \left(1+\frac{13m_W^2}{m_\chi^2}+\frac{m_W^4}{m_\chi^4}\right)\nonumber \\
&&+\left(1-\frac{c_H}{c_V}\right)^2 \left\{1+\frac{3m_W^2}{m_\chi^2}+\frac{31}{8}\frac{m_W^4}{m_\chi^4}
-\frac{3m_W^4}{m_G^2 m_\chi^2}+\frac{6m_W^4}{m_G^4}
\right\}
\Bigg] \left(1-\frac{m_W^2}{m_\chi^2}\right)^\frac{1}{2} .
\eea
For $c_H=c_V$, the above annihilation cross sections become
\bea
(\sigma \vrel)_{\chi{\bar\chi}\rightarrow ZZ}&\simeq&\vrel^2\cdot\frac{c^2_\chi c^2_V}{144\pi\Lambda^4}\frac{m^6_\chi}{(4m^2_\chi-m^2_G)^2+\Gamma^2_G m^2_G} \left(13+\frac{14m^2_Z}{m^2_\chi}+\frac{3m^4_Z}{m^4_\chi}\right)\left(1-\frac{m^2_Z}{m^2_\chi}\right)^\frac{1}{2}, \\
(\sigma v)_{\chi{\bar\chi}\rightarrow WW}&\simeq &\vrel^2\cdot \frac{c^2_\chi c^2_V}{72\pi\Lambda^4}\frac{m^6_\chi}{(4m^2_\chi-m^2_G)^2+\Gamma^2_G m^2_G}  \left(13+\frac{14m^2_W}{m^2_\chi}+\frac{3m^4_W}{m^4_\chi}\right)\left(1-\frac{m^2_W}{m^2_\chi}\right)^\frac{1}{2}.
\eea
For $c_H\gg c_V$, the annihilation cross sections for a pair of massive gauge bosons are
\bea
(\sigma \vrel)_{\chi{\bar\chi}\rightarrow ZZ}&\simeq &\frac{\vrel^2}{144\pi\Lambda^4}\frac{c^2_\chi c^2_Hm^6_\chi}{(4m^2_\chi-m^2_G)^2+\Gamma^2_G m^2_G} 
\left(1+\frac{3m_Z^2}{m_\chi^2}+\frac{31}{8}\frac{m_Z^4}{m_\chi^4}
-\frac{3m_Z^4}{m_G^2 m_\chi^2}+\frac{6m_Z^4}{m_G^4}
\right)\left(1-\frac{m_Z^2}{m_\chi^2}\right)^\frac{1}{2},~\\
(\sigma \vrel)_{\chi{\bar\chi}\rightarrow WW}&\simeq & \frac{\vrel^2}{72\pi\Lambda^4}
\frac{c^2_\chi c^2_H m^6_\chi}{(4m^2_\chi-m^2_G)^2+\Gamma^2_G m^2_G}
 \left(1+\frac{3m_W^2}{m_\chi^2}+
 \frac{31}{8}\frac{m_W^4}{m_\chi^4}
-\frac{3m_W^4}{m_G^2 m_\chi^2}+\frac{6m_W^4}{m_G^4}
\right) \left(1-\frac{m_W^2}{m_\chi^2}\right)^\frac{1}{2}. \qquad\phantom{\,}
\eea
On the other hand, for $c_H=0$, we obtain the annihilation cross sections into a pair of massless gauge bosons as
\bea
(\sigma \vrel)_{\chi{\bar\chi}\rightarrow \gamma\gamma}&\simeq& \vrel^2\cdot \frac{c^2_\chi c^2_\gamma}{12\pi\Lambda^4}\frac{m^6_\chi}{(4m^2_\chi-m^2_G)^2+\Gamma^2_G m^2_G},\\
(\sigma \vrel)_{\chi{\bar\chi}\rightarrow gg}&\simeq&  \vrel^2\cdot\frac{2c^2_\chi c^2_g}{3\pi\Lambda^4}\frac{m^6_\chi}{(4m^2_\chi-m^2_G)^2+\Gamma^2_G m^2_G}.
\eea
The annihilation cross section for $\chi{\bar\chi}\rightarrow \psi{\bar\psi}$ is
\bea
(\sigma \vrel)_{\chi{\bar\chi}\rightarrow \psi{\bar\psi}} \simeq \vrel^2 \cdot \frac{c^2_\chi c^2_\psi }{72\pi\Lambda^4}
\frac{m^6_\chi}{(4m^2_\chi-m^2_G)^2+\Gamma^2_G m^2_G} \left(1-\frac{m^2_h}{m^2_\chi}\right)^\frac{3}{2} 
\left(3+\frac{2m^2_\psi}{m^2_\chi}\right).
\eea

\def\theequation{E.\arabic{equation}}

\setcounter{equation}{0}

\vskip0.8cm
\noindent
{\Large \bf Appendix E:  DM annihilation cross sections in vector dark matter case} 
\vskip0.4cm
\noindent

In this section, we present the results of the annihilation cross sections for vector dark matter.

The annihilation cross section is
\bea
(\sigma \vrel)_{XX\rightarrow hh}\simeq \frac{2c^2_X c^2_H}{27\pi \Lambda^4}
\frac{m^6_X}{(4m^2_X-m^2_G)^2+\Gamma^2_G m^2_G}\left(1-\frac{m^2_h}{m^2_X}\right)^\frac{5}{2}.
\eea
Thus, the resulting annihilation cross section is s-wave.

The annihilation cross sections for vector dark matter going into a pair of massive gauge bosons, $XX\rightarrow AA$, are
\bea
(\sigma \vrel)_{XX\rightarrow ZZ} &\simeq&  \frac{2 c_{X}^2 c_V^2}{27\pi \Lambda^4} 
\frac{m_{X}^6}{(m_G^2-4 m_{X}^2)^2+\Gamma_G^2 m_G^2} 
\Bigg[\left(13+\frac{14m_Z^2}{m_{X}^2}+\frac{3m_Z^4}{m_{X}^4}\right)
-2  \left(1-\frac{c_H}{c_V}\right) \left(1+\frac{13m_Z^2}{m_{X}^2}+\frac{m_Z^4}{m_{X}^4}\right)\nonumber \\
&&+ \left(1-\frac{c_H}{c_V}\right)^2 \left\{1+\frac{3m_Z^2}{m_{X}^2}+\frac{115}{32}\frac{m_Z^4}{m_{X}^4}
-\frac{3}{4}\frac{m_Z^4}{m_G^2 m_{X}^2}+\frac{3}{2}\frac{m_Z^4}{m_G^4}
\right\}
\Bigg] \left(1-\frac{m_Z^2}{m_{X}^2}\right)^\frac{1}{2} , \\
(\sigma \vrel)_{XX\rightarrow WW} &\simeq&  \frac{4 c_{X}^2 c_V^2}{27\pi \Lambda^4} 
\frac{m_{X}^6}{(m_G^2-4 m_{X}^2)^2+\Gamma_G^2 m_G^2} 
\Bigg[\left(13+\frac{14m_W^2}{m_{X}^2}+\frac{3m_W^4}{m_{X}^4}\right)
-2  \left(1-\frac{c_H}{c_V}\right) \left(1+\frac{13m_W^2}{m_{X}^2}+\frac{m_W^4}{m_{X}^4}\right)\nonumber \\
&&+ \left(1-\frac{c_H}{c_V}\right)^2 \left\{1+\frac{3m_W^2}{m_{X}^2}+\frac{115}{32}\frac{m_W^4}{m_{X}^4}
-\frac{3}{4}\frac{m_W^4}{m_G^2 m_{X}^2}+\frac{3}{2}\frac{m_W^4}{m_G^4}
\right\}
\Bigg] \left(1-\frac{m_W^2}{m_{X}^2}\right)^\frac{1}{2}  .
\eea 
For $c_H=c_V$, the above annihilation cross sections become
\bea
(\sigma \vrel)_{XX\rightarrow ZZ} &\simeq&  \frac{2 c_{X}^2 c_V^2}{27\pi \Lambda^4} 
\frac{m_{X}^6}{(m_G^2-4 m_{X}^2)^2+\Gamma_G^2 m_G^2} 
\left(13+\frac{14m_Z^2}{m_{X}^2}+\frac{3m_Z^4}{m_{X}^4}\right)
 \left(1-\frac{m_Z^2}{m_{X}^2}\right)^\frac{1}{2} , \\
(\sigma \vrel)_{XX\rightarrow WW} &\simeq&  \frac{4 c_{X}^2 c_V^2}{27\pi \Lambda^4} 
\frac{m_{X}^6}{(m_G^2-4 m_{X}^2)^2+\Gamma_G^2 m_G^2} 
\left(13+\frac{14m_W^2}{m_{X}^2}+\frac{3m_W^4}{m_{X}^4}\right) \left(1-\frac{m_W^2}{m_{X}^2}\right)^\frac{1}{2}  .
\eea 
For $c_H\gg c_V$, the annihilation cross sections for a pair of massive gauge bosons are
\bea
(\sigma \vrel)_{XX\rightarrow ZZ} &\simeq&  \frac{2 c_{X}^2 c_H^2}{27\pi \Lambda^4} 
\frac{m_{X}^6}{(m_G^2-4 m_{X}^2)^2+\Gamma_G^2 m_G^2} 
\left(1+\frac{3m_Z^2}{m_{X}^2}+\frac{115}{32}\frac{m_Z^4}{m_{X}^4}
-\frac{3}{4}\frac{m_Z^4}{m_G^2 m_{X}^2}+\frac{3}{2}\frac{m_Z^4}{m_G^4}
\right) \left(1-\frac{m_Z^2}{m_{X}^2}\right)^\frac{1}{2} , \\
(\sigma \vrel)_{XX\rightarrow WW} &\simeq&  \frac{4 c_{X}^2 c_V^2}{27\pi \Lambda^4} 
\frac{m_{X}^6}{(m_G^2-4 m_{X}^2)^2+\Gamma_G^2 m_G^2} 
 \left(1+\frac{3m_W^2}{m_{X}^2}+\frac{115}{32}\frac{m_W^4}{m_{X}^4}
-\frac{3}{4}\frac{m_W^4}{m_G^2 m_{X}^2}+\frac{3}{2}\frac{m_W^4}{m_G^4}
\right)
 \left(1-\frac{m_W^2}{m_{X}^2}\right)^\frac{1}{2} .~~~~~~~~
\eea 
On the other hand, for $c_H=0$, we obtain the annihilation cross sections into a pair of massless gauge bosons as
\bea
(\sigma v)_{XX\rightarrow \gamma\gamma}&=&\frac{8c^2_X c^2_\gamma}{9\pi\Lambda^4}\frac{m^6_X}{(4m^2_X-m^2_G)^2+\Gamma^2_G m^2_G}, \\
(\sigma v)_{XX\rightarrow gg}&=&\frac{64c^2_X c^2_g}{9\pi\Lambda^4}\frac{m^6_X}{(4m^2_X-m^2_G)^2+\Gamma^2_G m^2_G}.
\eea
The annihilation cross section for $XX\rightarrow \psi{\bar\psi}$ is
\bea
(\sigma \vrel)_{XX\rightarrow \psi{\bar\psi}}&\simeq& \frac{4c^2_X c^2_\psi }{27\pi \Lambda^4}
 \frac{m^6_X}{(4m^2_X-m^2_G)^2+\Gamma^2_G m^2_G}\left(3+\frac{2m^2_\psi}{m^2_X}\right)\left(1-\frac{m^2_\psi}{m^2_X}\right)^\frac{3}{2}.
\eea

\def\theequation{F.\arabic{equation}}

\setcounter{equation}{0}

\vskip0.8cm
\noindent
{\Large \bf Appendix F:  DM annihilation cross sections when $m_{DM} >  m_G$} 
\vskip0.4cm
\noindent
 
In this case, $t$ and $u$ channels for the dark matter annihilation will open, and the all of these are S-wave as following results
in the limit of the width of a dark matter can be negligible compared to a dark matter mass, 
\bea
(\sigma \vrel)_{SS\rightarrow G,G} &\simeq& \frac{4 c_{S}^4 m_{S}^2}{9 \pi \Lambda^4 }
\frac{(1-r)^\frac{9}{2}}{r^4  (2-r)^2} , \\
(\sigma \vrel)_{\chi \bar\chi \rightarrow G,G} &\simeq& \frac{c_{\chi}^4 m_{\chi}^2}{16 \pi \Lambda^4 }
\frac{(1-r)^\frac{7}{2}}{r^4 (2-r)^2} , \\
(\sigma \vrel)_{X_\mu X_\nu \rightarrow G,G} &\simeq&
\frac{c_{X}^4 m_{X}^2}{324 \pi \Lambda^4 } 
\frac{\sqrt{1-r}}{r^4  (2-r)^2} \,
\bigg(176+192 r+1404 r^2-3108 r^3 \nonumber \\
&&+1105 r^4+362 r^5+34 r^6 \bigg)
\eea
with $r = \left(\frac{m_G}{m_{DM}}\right)^2$.  

\def\theequation{G.\arabic{equation}}

\setcounter{equation}{0}

\vskip0.8cm
\noindent
{\Large \bf Appendix G:  DM annihilation cross sections with a radion mediator} 
\vskip0.4cm
\noindent

In this section, we present annihilation cross sections of dark matters into higgs pair through a radian mediator.
\bea
(\sigma \vrel)_{SS\rightarrow h,h} &\simeq&\frac{(c_H^r c_S^r)^2}{16\pi\Lambda^4}\frac{ m_S^6}{(m_R^2-4 m_S^2)^2+\Gamma_R^2 m_R^2}
\left(2+\frac{m_h^2}{m_S^2}\right)^2 \left(1-\frac{m_h^2}{m_S^2}\right)^\frac{1}{2} , \\
(\sigma \vrel)_{\chi \bar\chi \rightarrow h,h} &\simeq&  \vrel^2 \cdot \frac{(c_H^r c_\chi^r)^2}{1152\pi\Lambda^4}\frac{ m_S^6}{(m_R^2-4 m_\chi^2)^2+\Gamma_R^2 m_R^2}
\left(2+\frac{m_h^2}{m_\chi^2}\right)^2 \left(1-\frac{m_h^2}{m_\chi^2}\right)^\frac{1}{2}  \\
(\sigma \vrel)_{X_\mu X_\nu \rightarrow h,h} &\simeq&
 \frac{(c_H^r c_{X}^r)^2}{432\pi\Lambda^4}\frac{ m_{X}^6}{(m_R^2-4 m_{X}^2)^2+\Gamma_R^2 m_R^2}
\left(2+\frac{m_h^2}{m_{X}^2}\right)^2 \left(1-\frac{m_h^2}{m_{X}^2}\right)^\frac{1}{2} , 
\eea
As we can see, scalar dark matter and vector dark matter case, annihilation is $S$-wave while in a fermion dark matter case, the annihilation is $P$-wave suppressed.

 \def\theequation{G.\arabic{equation}}

\setcounter{equation}{0}

\vskip0.8cm
\noindent
{\Large \bf Appendix G:  The sum of Kaluza-Klein graviton modes } 
\vskip0.4cm
\noindent

We consider the sum of KK modes for the s-channel process with KK graviton exchanges such as $XX\rightarrow G^{(n)}\rightarrow {\rm SM}\,\,{\rm SM}$, which is the annihilation of dark matter $X$.
The amplitude of the process is given by
\be
{\cal M}={\cal A} {\cal S}
\ee
where ${\cal A}$ is the matrix element corresponding to the interactions of KK gravitons to dark matter and SM particles and ${\cal S}$ is given by the sum of KK graviton propagators, 
\be
{\cal S}(s)=\frac{1}{\Lambda^2}\sum_{n=1}^\infty  \frac{1}{s-m^2_n+i \,m_n \Gamma_n}. \label{kksum0}
\ee
Here, $\Gamma_n$ denotes the total width of the graviton with KK number $s$ and mass $m_n$ and is given by
\be
\Gamma_n\approx \eta \,m_n \left(\frac{m_n}{\Lambda}\right)^2,\quad\quad \eta=\frac{c^2_H}{240\pi}. 
\ee
The KK graviton masses are determined by the zeros of $J_1(x_n)$ as $m_n=x_n k \Lambda /M_P$, with $x_n=\pi(n+1/4)+{\cal O}(n^{-1})$.

In order to perform the KK sum, we rewrite eq.~(\ref{kksum0}) as
\be
{\cal S}(s)=\sum_{i=1}^\infty \frac{1}{ax^4_n-bx^2_n+c}=\frac{1}{a(\sigma^2-\rho^2)}\sum_{n=1}^\infty \bigg(\frac{1}{x^2_n -\sigma^2}-\frac{1}{x^2_n-\rho^2}\bigg)
\ee
with
\be
a=i\eta\, \Big(\frac{k}{M_P}\Big)^4 \Lambda^4,\quad b=\Big(\frac{k}{M_P}\Big)^2 \Lambda^4,\quad c=s\Lambda^2,
\ee
and 
\bea
\sigma^2&=&\frac{s}{\Lambda^2}\left(\frac{M_P}{k}\right)^2\,\frac{2}{1+\sqrt{1-4i\eta\,\frac{s}{\Lambda^2}}}, \\
\rho^2&=& \frac{1}{2i\eta} \left(\frac{M_P}{k}\right)^2 \left(1+\sqrt{1-4i\eta\,\frac{s}{\Lambda^2}}\right).
\eea
Using the following formula,
\be
\sum_{n=1}^\infty \frac{1}{x^2_n-\sigma^2}=\frac{1}{2\sigma}\,\frac{J_2(\sigma)}{J_1(\sigma)},
\ee
we obtain \cite{kksum}
\bea
{\cal S}(s)=\frac{1}{2a(\sigma^2-\rho^2)}\left(\frac{1}{\sigma}\,\frac{J_2(\sigma)}{J_1(\sigma)} -\frac{1}{\rho}\,\frac{J_2(\rho)}{J_1(\rho)}\right).
\eea

Now, for $\eta s\ll \Lambda^2$, we take the approximate forms,
\bea
\sigma &\simeq& \frac{x_1\sqrt{s}}{m_1}\left(1+\frac{i\eta}{2}  \frac{s}{\Lambda^2} \right), \\
|\rho| &\simeq& \frac{1}{\sqrt{\eta}} \,\frac{x_1\Lambda}{m_1} \gg |\sigma|
\eea
where use is made of $m_1=x_1 k\Lambda/M_P$ with $x_1=3.83$.
In this case, the KK sum becomes
\bea
{\cal S}(s)\simeq -\frac{1}{4\Lambda^2\sqrt{s}}\,\frac{x_1}{m_1}\,\frac{J_2(\sigma)}{J_1(\sigma)}.
\eea

\end{document}